\documentclass[%
 reprint,
nofootinbib,
 amsmath,amssymb,
 aps,
 prd,
]{revtex4-1}
\pdfoutput=1
\usepackage{graphicx}
\usepackage{dcolumn}

\usepackage[utf8]{inputenc} 
\usepackage{hyperref}
\usepackage{color}
\usepackage{soul}
\usepackage{ulem}
\usepackage{cancel}

\newcommand{\be}{\begin{equation}}
\newcommand{\ee}{\end{equation}}
\newcommand{\ba}{\begin{eqnarray}}
\newcommand{\ea}{\end{eqnarray}}

\begin{document}

\preprint{INR-TH}

\title{Dark matter component decaying after recombination: \\ 
Sensitivity to BAO and RSD probes}

\author{A. Chudaykin}
 \email{chudy@ms2.inr.ac.ru}
\affiliation{Institute for Nuclear Research of the Russian Academy of Sciences,
  Moscow 117312, Russia}%
\affiliation{Moscow Institute of Physics and Technology, 
  Dolgoprudny 141700, Russia}%

\author{D. Gorbunov}
 \email{gorby@ms2.inr.ac.ru}
\affiliation{Institute for Nuclear Research of the Russian Academy of Sciences,
  Moscow 117312, Russia}%
\affiliation{Moscow Institute of Physics and Technology, 
  Dolgoprudny 141700, Russia}%

\author{I. Tkachev}
 \email{tkachev@ms2.inr.ac.ru}
\affiliation{Institute for Nuclear Research of the Russian Academy of Sciences,
  Moscow 117312, Russia}%
\affiliation{Novosibirsk State University,
  Novosibirsk 630090, Russia}%

\begin{abstract}

It has been recently suggested~\cite{Berezhiani:2015yta} that a
subdominant fraction of dark matter decaying after recombination may
alleviate tension between high-redshift (CMB anisotropy) and
low-redshift (Hubble constant, cluster counts) measurements. In this
report, we continue our previous study~\cite{Chudaykin:2016yfk} of the
decaying dark matter (DDM) model adding all available recent baryon
acoustic oscillation (BAO) and redshift space distortions (RSD)
measurements. We find, that the BAO/RSD measurements generically
prefer the standard $\Lambda$CDM and combined with other cosmological
measurements impose an upper limit on the
DDM fraction at the level of
$\sim$\,5\,\%, strengthening by a factor of 1.5 limits obtained in 
\cite{Chudaykin:2016yfk} mostly from CMB data. However,
the numbers vary from one analysis to other based on
the same Baryon Oscillation Spectroscopic Survey (BOSS) Data Release
12 (DR12) galaxy sample. Overall, the model with a few percent
DDM fraction provides a better fit to the combined cosmological data
as compared to the $\Lambda$CDM: the cluster counting and direct
measurements of the Hubble parameter are responsible for that. 
The improvement can be as large as 1.5\,$\sigma$ and
grows to  3.3\,$\sigma$ when the CMB lensing power amplitude ${\rm
  A_L}$ is introduced  as a free fitting parameter. 

\end{abstract}

\maketitle

\section{Introduction\label{sec:intro}}

It is known that an additional form of matter which clusters
gravitationally but otherwise is (almost) immune to other interactions
is needed to describe cosmological data. The corresponding matter fraction
is called dark matter (DM), and its nature remains elusive so
far. Moreover, it is unknown whether DM is one-component or, in turn,
consists of several different species. The latter
idea developed in the 1980s
\cite{Flores:1986jn,Doroshkevich:1989bf} and recently gained renewed
interest because of both the growth in precision of cosmological
measurements and the appearance of a tension between low-redshift
measurements and predictions of the standard $\Lambda$CDM cosmology
based on the high-redshift observations.

In Ref.~\cite{Berezhiani:2015yta} it was argued that the subdominant DM
component decaying after recombination may alleviate the 
  aforementioned tension~\cite{Ade:2013zuv}.  Instead of fitting of
this model to the Planck data, it was simply assumed that all
cosmological parameters at recombination correspond to the Planck
derived values. This is reasonable since the  assumed unstable DM
fraction decays after recombination. However, at the level of modern
precision achieved in cosmological data analyses, this simple approach is not sufficient anymore,
since CMB anisotropies are subject to the gravitational lensing at a
later epoch. This effect is observable with Planck and should be
accounted for.
 
In the follow-up paper~\cite{Chudaykin:2016yfk}, a proper fitting of
the decaying DM model has been carried out. There, in addition to the Planck
likelihood for TT,TE,EE power spectra~\cite{Ade:2015xua}, the direct
measurement of Hubble constant $H_0$~\cite{Riess:2011yx} and probes of
matter clustering $\sigma_8$ and matter fraction in present energy
density $\Omega_m$ from the Planck cluster counts~\cite{Ade:2015fva}
have been considered.  It was found that the model with decaying dark
matter (DDM) is indeed somewhat more preferable in comparison with
base $\Lambda$CDM: however, the fraction of DDM is away off the
original suggestion\,\cite{Berezhiani:2015yta} being severely
restricted by lensing observed in TT spectrum. Notably, a final
verdict turns out to be highly dependent upon the choice of
  additional data set: polarization at low multipoles or direct
Planck probes of lensing power spectrum.

At the same time, a whole layer of precise
measurements of baryon acoustic oscillation (BAO) and more complicated
probes of redshift space distortions (RSD) which may shed light on the
nature of the dark sector has not been studied
in~\cite{Chudaykin:2016yfk}. Beside  
low-statistics measurements at low redshifts (which are generally
consistent with the $\Lambda$CDM framework~\cite{Ade:2015xua}), there
are a number of rather precise  middle-redshift probes that differ   
in type of constraints, redshift separation, sample volume
and analysis procedure. These probes are known to be not so united in
exploring the $\Lambda$CDM: in some cases the best-fit values of
the cosmological parameters noticeably deviate. 
For instance, the BAO or RSD signal can be
extracted from correlation function in configuration space or from
power spectrum in Fourier space, sample volumes corresponding to
different redshifts may be independent or overlap. The first goal of the
current research is to explore the cosmological implication of various
BAO and RSD measurements based on the BOSS Data Release 12 (DR12) galaxy
sample in the range $0.15<z<0.75$ in combination with other
cosmological data to observe possible hints of DDM.

More recently, the BAO signal has also been extracted from
flux-transmission correlations in the Ly-$\alpha$ forest and from
cross-correlations of the Ly-$\alpha$ absorption sites with positions
of quasars. These measurements correspond to much higher effective
average redshift $z_{\rm eff}\approx2.3-2.4$ and therefore provide
independent probes of the Universe expansion at that
times. Remarkably, such high-redshift probes are in some tension with
the $\Lambda$CDM prediction.  If obtained constraints are not plagued
by  
unaccounted systematics, then  the BAO signals in the Ly-$\alpha$ forest
hint to cosmology beyond the $\Lambda$CDM pattern. Moreover, if the
statistical errors in these two measurements are almost completely
uncorrelated, cross- and autocorrelation analyses can be combined into
one set which leads to a more pronounced $\approx2.3\sigma$
discrepancy \cite{Bourboux:2017cbm} between BAO high-redshift probes
and the CMB-dominated best fit of $\Lambda$CDM.

It was already argued~\cite{Berezhiani:2015yta} that the model with
subdominant unstable fraction of DM decaying after recombination is
capable of easing the tension above. However in
Ref.~\cite{Berezhiani:2015yta} no proper likelihood of BAO at large
redshifts have been exploited. This lacuna is filled in the
present work.

It is worth mentioning that popular extensions of the base
$\Lambda$CDM model such as nonzero space curvature or varying in time
dark energy equation-of-state only partially ease the tension between
BAO measurements and are not able to eliminate the Ly-$\alpha$ anomaly
completely, see for example~\cite{Evslin:2015uwa,Evslin:2016gre}. For
this reason, such a prominent discrepancy between BAO probes at low
and high redshifts still deserves a special study. In the current
research, we explore this strong tension in light of the DDM
framework.

The final goal of the present paper is to classify the set of BAO and
RSD probes from the BOSS DR12 galaxy sample by their preferences for
DDM and find out whether an admixture of DDM really helps to reconcile
the Ly-$\alpha$ BAO anomaly with other cosmological observables.
In this analysis we also consider various 
up-to-date cosmological and astrophysical data similarly
to the Ref.~\cite{Chudaykin:2016yfk} and obtain actual constraints on the
parameters of the DDM model.
    
The paper is organized as follows. In Sec.\,\ref{sec:model}, we
describe the cosmological model, the cosmological data sets  utilized, and
the numerical procedure adopted to explore  model parameter
space. We present the obtained constraints on DDM in
Sec.\,\ref{sec:constraints} and summarize our results and discuss
future prospects in Sec.\,\ref{sec:conclusions}.  


\section{The model, data sets and procedure}
\label{sec:model}

\subsection{Decaying dark matter model}
\label{subsec:DDM}

We assume that dark matter consists of stable $\Omega_{sdm}$ and
decaying $\Omega_{ddm}$ parts, and the unstable particles decay into
dark radiation (e.g. unknown ultrarelativistic particles).  Following
Ref.~\cite{Berezhiani:2015yta} we define a fraction of the decaying
part, $F$, in terms of initial densities $\omega_{i}\equiv\Omega_ih^2$
as $F\equiv{\omega_{ddm}}/({\omega_{sdm}+\omega_{ddm}})$. Here the
present value of the Hubble parameter is parametrized as $H_0\equiv
h\times 100$\,km/s/Mpc.  We measure the width of corresponding decay,
$\Gamma$, in units of km/s/Mpc adopted for the Hubble parameter
determining the Universe expansion rate. Following
Ref.~\cite{Berezhiani:2015yta} we assume also that the DM decays
mainly after recombination epoch.  Then the CMB spectra at last
scattering are intact and the primary cosmological parameters are
close to the Planck derived values. This implies constraint
$\Gamma<5000$\,km/s/Mpc.

\subsection{Cosmological data sets}
\label{subsec:data}

\subsubsection{Planck, Hubble and  cluster counts}
We employ the full Planck likelihood for TT,TE,EE power spectra at
multipoles $l>30$~\cite{Ade:2015xua} to account for the lensing effects of
CMB anisotropies properly as explained
in~\cite{Chudaykin:2016yfk}. We also exploit measurements of
polarization at low multipoles~\cite{Ade:2015xua} and direct probes of
the lensing power spectrum $C_{l}^{\phi\phi}$ calculated from the
non-Gaussian parts of 15 different 4-point
functions~\cite{Ade:2015len}. Using the two latter probes
simultaneously imposes the tightest constraint on the parameter $F$
according to~\cite{Chudaykin:2016yfk}.
We do not use the last Planck polarization
constraint~\cite{Adam:2016hgk} here because a proper likelihood is not
available yet.

For the low-redshift cosmological probes, we take the galaxy cluster
counts from Planck catalogues~\cite{Ade:2015fva} as in
Ref.~\cite{Chudaykin:2016yfk} and a more recent and precise direct
measurement of
the Hubble constant~\cite{Riess:2016jrr}. These data sets are 
conflicting currently with  the Planck high-redshift measurements and a nonzero
fraction of DDM may alleviate this tension as reported
in Refs.\,\cite{Berezhiani:2015yta,Chudaykin:2016yfk}.

We refer to the combined set of Planck, Hubble and cluster counts data
listed above as the ''Base'' data set.

\subsubsection{BAO  probes at low redhifts}

Lowest redshift data sets, which have inherent limited
statistics, provide only an isotropic measurement of the angle-average
distance ratio $D_V/r_d\approx D_H^{0.7}D_A^{0.3}/r_d$, where
$D_H(z)=c/H(z)$. In this paper, we consider the BAO signal from the
SDSS Main Galaxy Sample (MGS) at $z=0.15$~\cite{Ross:2014qpa} and the
Six-degree-Field Galaxy Survey (6dFGS) at
$z=0.106$~\cite{Beutler:2011hx}, and call them MGS and 6dFGS,
respectively. Such measurements have the insignificant overlapping
galaxy volume and can be considered as independent surveys. Moreover,
isotropic measurements at low redshifts and any other anisotropic BAO
probe at higher $z$ can also be treated as independent and
combined afterwards.

\subsubsection{BAO  probes at  middle redshifts}
\label{subsec:data_BAOlowz}

The most precise BAO measurements provide a universal ruler, the
comoving sound horizon at the baryon drag epoch $r_d$, which has been
used to measure the expansion of the Universe at different epochs. The
recent analyses of the BOSS DR12 galaxy
sample~\cite{Cuesta:2015mqa,Gil-Marin:2015nqa,Alam:2016hwk} have
satisfactory statistical power to measure the BAO peak position in
both the line-of-sight and transverse directions which imply
constraints on the angular diameter distance $D_A(z)$ and Hubble
parameter $H(z)$ simultaneously in the units of the standard ruler
$r_d$.  In this paper, we use the following BAO probes, 
all based on the latest BOSS Data Release 12.

{\bf i)} We utilize results of Refs.~\cite{Cuesta:2015mqa}
and~\cite{Gil-Marin:2015nqa} where two distinct samples of luminous
galaxies in the ranges $0.15<z<0.43$ and $0.43<z<0.7$ are used to
extract the BAO signal from the moments of Fourier-space power
spectrum or correlation functions, respectively. We also exploit one
consensus BOSS result~\cite{Alam:2016hwk} which consists of various
similar to each other BAO probes and where the overall sample is divided
into three bins in the redshift space.

{\bf ii)} There is a way to capture a tomographic (continuous)
redshift-evolution of the BAO scale. For that, the overall galaxy
sample should be divided into a large number of overlapping redshift
bins to perform the correlated BAO analysis in each slice. If this
procedure ensures sufficient galaxy counts in each subsample, it
provides reliable measurements in each redshift bin. The tomographic
technics is used to find proper constraints on time-evolving quantities 
such as the dark energy equation of state, $\omega(z)$. 
However, such measurements may be of interest with regard to
constraints on DDM for another reason. Tomographic analyses of BAO 
provide the largest number of quite solid measurements 
which implies the highest statistical weight and
brings the greatest contribution to the $\chi^2$ function among
  other BAO probes. Since a tomographic 
  analysis employs overlapping sample volumes in different redshift bins, 
  this procedure requires a proper use of the full covariance
matrix. For such a ''tomographic'' probe, we use recent results of Refs.~\cite{Zhao:2016das,Wang:2016wjr}. The
probe is based on the power spectrum and correlation function reconstruction
technics, respectively, and traces the BAO signal in the range
$0.2<z<0.75$ in nine overlapping redshift slices.

\smallskip

Generally, the main advantage of BAO probes rests in their pure
geometrical character. They are not affected by uncertainties in
nonlinear evolution of the matter density field and, therefore, can
impose very robust constraints on model parameters. 

\subsubsection{RSD probes at middle redshifts}
\label{subsec:data_RSDlowz}

The RSD anisotropy caused by peculiar velocities of baryons and dark
matter and observed in multipole moments of the galaxy power spectrum
and two-point correlation functions is a powerful tool for
constraining the growth rate of structures in the Universe. Transverse
versus line-of-sight anisotropies in the redshift space can be
approximated in the linear theory by the density-velocity correlation
power spectrum, so the RSD tests the normalized growth rate,
$f(z)\sigma_8^{~}(z)={\sigma_8^{(vd)}(z)^2}/{\sigma_8^{~}(z)}$, where
$\sigma_8^{(vd)}$ is the smoothed density-velocity correlation
averaged over $8h^{-1}$ Mpc~\cite{Ade:2015xua}. Unfortunately, these
measurements are affected by nonlinearities on small scales, galaxy
bias and harmful degeneracies with background parameters. For this
reason, constraints imposed by RSD tests are significantly looser than
those obtained in BAO analyses. Nevertheless, the RSD approach
provides an additional independent probe of the cross-correlation between
the LSS and the velocity anisotropy in the Universe at different
times, which may be critical when testing DDM predictions.

For our purpose we exploit single-probe
measurements~\cite{Chuang:2016uuz} from the BOSS DR12 galaxy
sample split into two and four redshift bins where the RSD signal
has been extracted from correlation functions. We also employ two
analyses from~\cite{Gil-Marin:2015sqa} based on different mock
realizations with two redshift bins and where multipole moments of the
galaxy power spectrum were exploited instead.  Finally, we investigate
consensus constraints on $D_A(z)$, $H(z)$ and $f\sigma_8(z)$ at three
effective redshifts obtained in RSD survey~\cite{Alam:2016hwk} as
well.

\subsubsection{BAO Ly-$\alpha$}
\label{subsec:data_Lya}

It was proposed that the Lyman-$\alpha$ forest of absorption of light
from quasars can be used to trace the underlying matter density field,
so the BAO signal at higher redshifts may be found there. The first
such signal was detected in the cross-correlations between the
Ly-$\alpha$ forest absorption and the distribution of quasars using
the DR11 sample~\cite{Font-Ribera:2013wce}. A little bit later similar
signal was found in the Ly-$\alpha$ forest autocorrelation function in
the DR11 study of~\cite{Delubac:2014aqe}. Both measurements are in
some tension with predictions of the flat $\Lambda$CDM cosmological
model. Moreover, it was argued in~\cite{Delubac:2014aqe} that
statistical errors in these two BAO probes are uncorrelated and one
can treat them as independent surveys. In this case, the tension
between combined cross- and autocorrelation BAO measurements and the
$\Lambda$CDM-Planck best fit cosmology becomes even stronger and
reaches the level of $\approx2.5\sigma$~\cite{Delubac:2014aqe}.

In our study, we exploit the state-of-the-art
cross-correlation~\cite{Bourboux:2017cbm} and
autocorrelation~\cite{Bautista:2017zgn} measurements based on the
latest BOSS Data Release 12 where several improvements in the analysis
procedure were also developed. Exploiting these two kinds of
measurements simultaneously gives us the largest mismatch
between BAO probes at high redshifts and the $\Lambda$CDM
prediction. Using the first author's names we denote the corresponding
data sets as Bourboux and Bautista, respectively.

The $15\%$ increase of the sample volume in the cross-correlation and
autocorrelation analyses~\cite{Bourboux:2017cbm,Bautista:2017zgn}
over the previous studies~\cite{Font-Ribera:2013wce,Delubac:2014aqe}
is mainly responsible for $0.5\sigma$ reduction in $D_H(2.34)/r_s$
relative to the previous measurement based on the BOSS DR11 quasar
sample as argued in~\cite{Delubac:2014aqe}. It reduces mismatch
between combined BAO probes at high redshifts and the
$\Lambda$CDM-Planck best fit cosmology to the level of
$\approx2.3\sigma$~\cite{Bourboux:2017cbm} which alleviates the
Ly-$\alpha$ anomaly insignificantly.


\subsection{Numerical procedure}

We test two-component DDM model against data using the Markov chain
Monte Carlo (MCMC) approach within the \textsc{Monte Python}
package~\cite{Audren:2012wb}.  We modified the CLASS \textsc{Bolzman}
code~\cite{Lesgourgues:2011re,Blas:2011rf} to implement computation of
$f\sigma_8$ at different redshifts so that RSD probes can be properly
adopted. In particular, calculation of the power spectrum $P^{(v
  d)}(k)$ has been added, where $d$ stands for the total matter
density fluctuations and $v =-\nabla{ \vec v}_N/H$, where ${\vec v_N}$
is peculiar velocity field of baryons and dark matter.

Eight free parameters are varied in the fitting procedure. Two of them
are inherent to the DDM model: the fraction $F$ and the width
$\Gamma$. Remaining six correspond to underlying $\Lambda$CDM
cosmology: the angular scale of the sound horizon $r_s$ at
last-scattering $\theta_*\equiv r_s(z_*)/D_A(z_*)$, the baryon density
$\omega_b $, initial CDM density $\omega_{cdm} = \omega_{sdm} +
\omega_{ddm}$, the optical depth $\tau$, the scalar spectral index
$n_s$, and the amplitude of the primordial power spectrum $A_s$. We
adopt spatially flat Universe and assume normal neutrino hierarchy
pattern with the total active mass $\sum m_\nu=0.06$\,eV.


\section{Constraints on DDM}
\label{sec:constraints}

\subsection{$\chi^2$ analysis of BAO and RSD probes at middle redshifts}
\label{subsec:chi}

\begin{figure}
\includegraphics[keepaspectratio,width=\columnwidth]{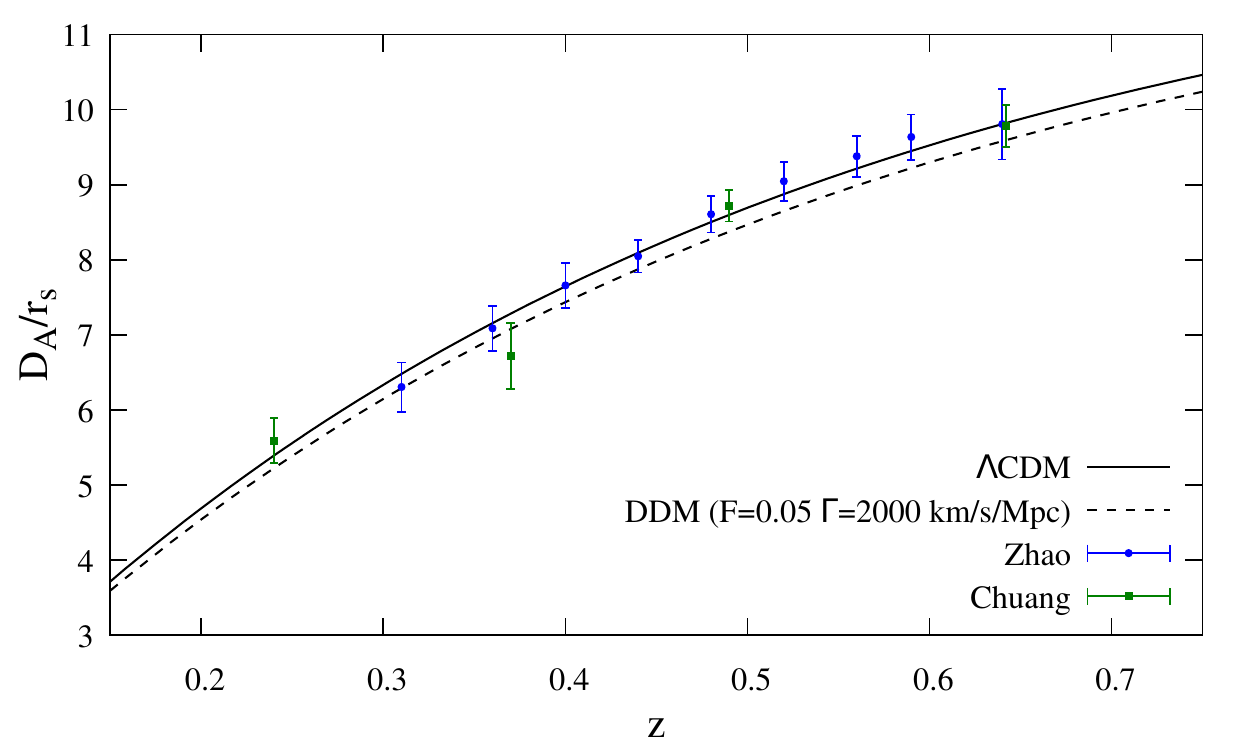}
\caption{The redshift evolution of $D_A/r_s$ in the $\Lambda$CDM model
  fitted to the Base data set (solid line) and in the DDM cosmology
  with reference values $F=0.05$ and $\Gamma=2000$ km/s/Mpc while 
  remaining six standard parameters are kept the same
  as in $\Lambda$CDM (dashed
  line). Zhao result is shown by blue boxes with error bars which show
  $\pm1\sigma$ uncertainties. Chuang likelihood is illustrated by
  green dots with error bars.}
\label{fig:DA} 
\end{figure}
\vspace{-2cm}
\begin{figure}[!t]
\includegraphics[keepaspectratio,width=\columnwidth]{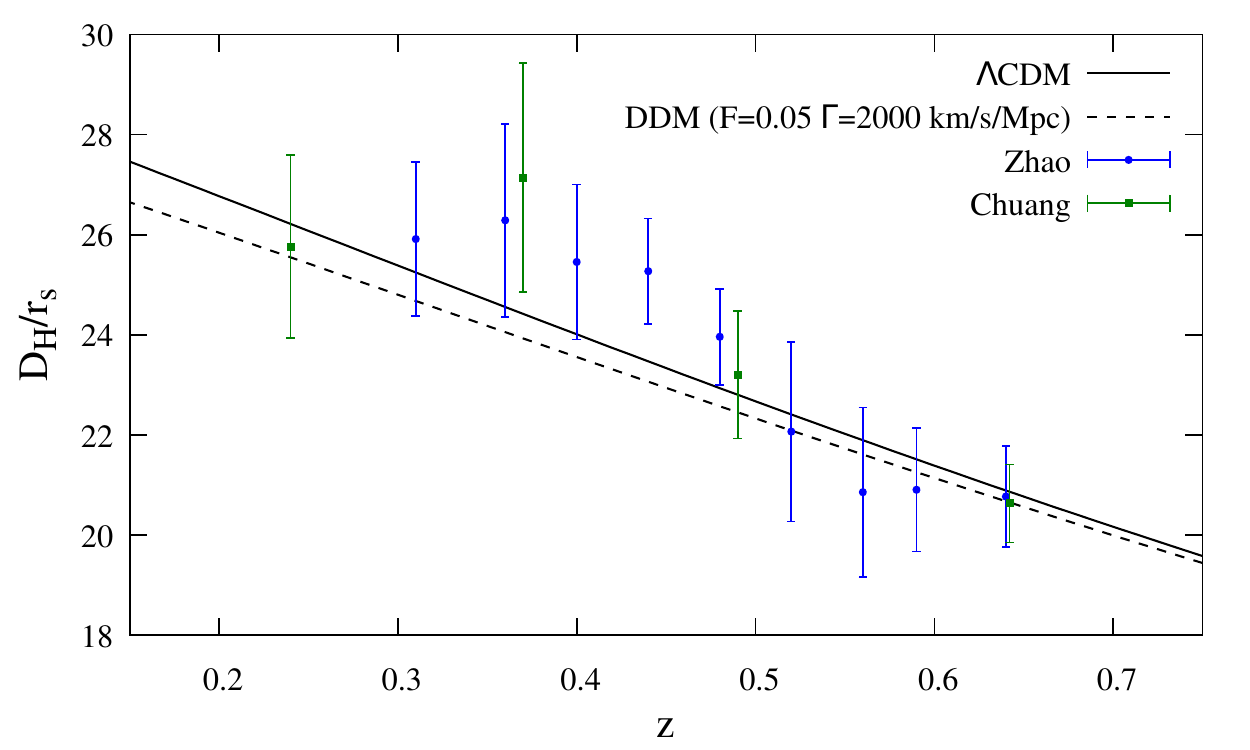}
\caption{\label{fig:DH} 
Same as Fig.~\ref{fig:DA} but for $D_H/r_s$ evolution.}
\end{figure}
\vspace{2cm}

We start with the preliminary $\chi^2$ analysis of each of the BAO and RSD probes at
middle redshifts from the BOSS DR12 galaxy sample in order to
find out which of them favors (or disfavors) DDM cosmology the most
and the least in comparison with standard $\Lambda$CDM. For that, we
utilize the corresponding covariant matrices available on the SDSS
website \href{http://www.sdss3.org/}{http://www.sdss3.ortrg/}.

According to our $\chi^2$ analysis we calculate for each data set and
different cosmologies the following quantity
\begin{equation}\label{chi2}
\chi^2=(x_{\rm data}-x_{\rm best-fit})^TC_{cov}^{-1}(x_{\rm data}-x_{\rm best-fit})
\end{equation} 
where $x_{\rm data}$ is the vector of mean values determined by
a particular BAO/RSD measurement, $x_{\rm best-fit}$ stands for the vector of
best-fit values obtained in the $\Lambda$CDM or DDM cosmology within
the Base data set and $C_{cov}$ denotes a covariant matrix of
corresponding measurements at different redshifts. Here we assume
  that other six base parameters are tightly constrained by the Base
  data set and all BAO/RSD probes have inefficient statistical weight
  to change them significantly. 

We find the $\chi^2$ value, Eq.~\eqref{chi2}, for various cosmologies and obtain the
difference $\Delta\chi^2=\chi^2_{\rm \Lambda CDM}-\chi^2_{\rm DDM}$
between $\Lambda$CDM and DDM patterns for five BAO and five RSD probes
from the same BOSS DR12 galaxy sample within $0.15<z<0.75$ 
and mentioned in subsec.~\ref{subsec:data_BAOlowz}
and~\ref{subsec:data_RSDlowz}. As a result of this procedure, we
select a tomographic probe based on the power spectrum reconstruction
technics, Ref.~\cite{Zhao:2016das}, as the one which prefers $\Lambda$CDM model
most strongly in comparison with DDM,  with $\Delta\chi^2=-8.14$. On the
other hand, {\it while sets which would favor DDM are absent,} the RSD
analysis based on the most recent single-probe reconstruction
technics~\cite{Chuang:2016uuz} with redshift interval split into
four parts 
shows $\Delta\chi^2=-2$
which provides the lowest inconsistency with DDM. Thus, we have two
cosmological probes from the same BOSS DR12 galaxy sample which
provide notably different constraints on the DDM model. We mark these
sets as Zhao and Chuang (again adopting the first authors'
names), and believe these sets embrace all other available BAO and RSD
probes in BOSS DR12.\footnote{Earlier  BAO studies based on previous
  LSS data releases  favored DDM a bit more, see Ref.\,\cite{Berezhiani:2015yta}.}. In what follows we select them for global fitting together
with the Base data set.

\begin{figure}[!t]
\includegraphics[keepaspectratio,width=\columnwidth]{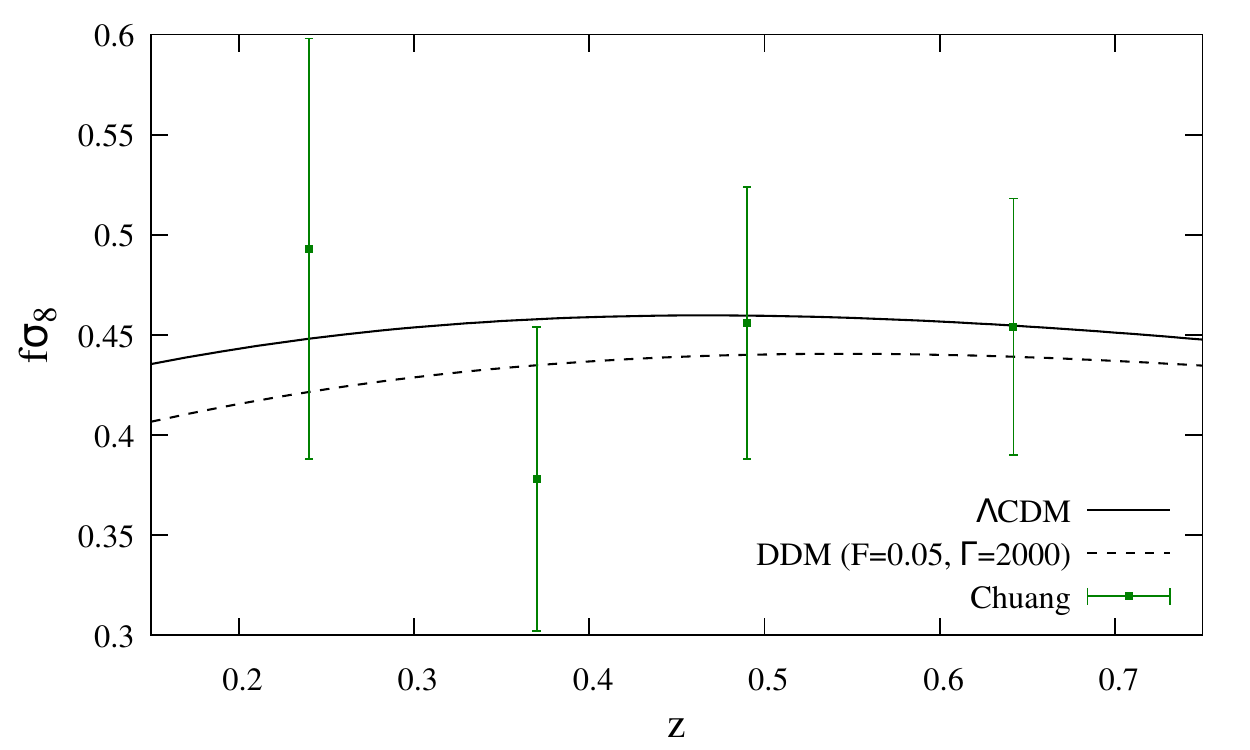}
\caption{\label{fig:fsigma8} 
Same as Fig.~\ref{fig:DA} but for $f\sigma_8$ evolution. Zhao data set is absent here because pure BAO measurements do not impose  constraints on $f\sigma_8$.}

\end{figure}
\vspace{-2cm}
\begin{figure}[!t]
\includegraphics[keepaspectratio,width=\columnwidth]{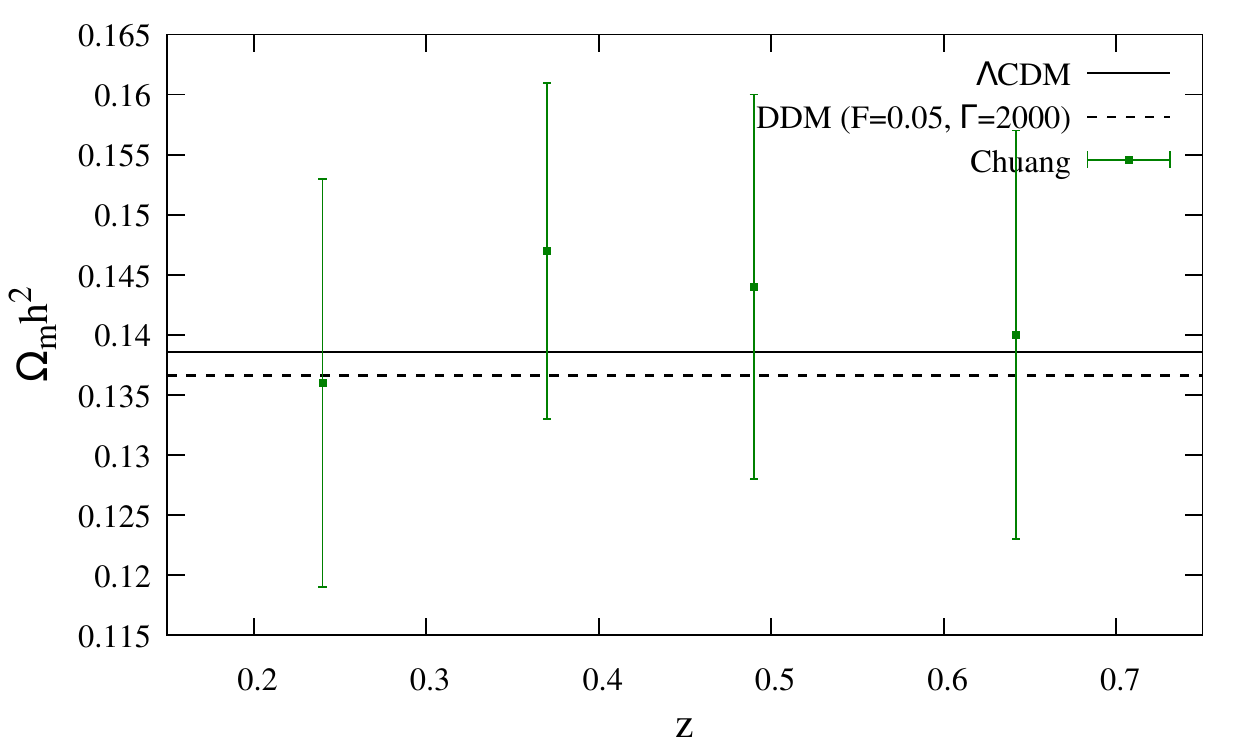}
\caption{\label{fig:OmegaM} 
Same as Fig.~\ref{fig:DA} but for $\omega_m=\Omega_m\,h^2$ evolution. Zhao data set does not constrain $\omega_m$.}
\end{figure}
\vspace{2cm}

Figures~\ref{fig:DA}, \ref{fig:DH}, \ref{fig:fsigma8}
and \ref{fig:OmegaM} illustrate the
behavior of DDM and $\Lambda$CDM with respect to selected
data sets. Figs.~\ref{fig:DA}, \ref{fig:DH} reveal comparable errors imposed by
Zhao and Chuang likelihoods on $D_A/r_s$ and $D_H/r_s$ parameters. Since the Zhao
probe contributes 18 measurements to the cosmological fit
whereas the Chuang set does only 4, the Zhao
data set possesses higher statistical weight with respect to $D_A/r_s$, $D_H/r_s$ and indeed can be responsible for the most robust constraints on model parameters.
Figures~\ref{fig:fsigma8} and \ref{fig:OmegaM} display 
measurements of $f\sigma_8$ and $\Omega_mh^2$. The Chuang likelihood exhibits there
too loose constraints on $f\sigma_8$ and $\Omega_mh^2$ which provide a subdominant contribution of these measurements 
to the $\chi^2$ function. So, the Zhao data set is really able to put the strongest 
constraints on the DDM model whereas the Chuang likelihood imposes only a 
mild constraint.
We warn the readers that depicted
pictures do not contain complete information about the measurements
and serve only for the  illustration since parameters are correlated at a
given redshift. Moreover, the measurements in the tomographic probe 
  at different redshifts are also correlated. All correlations are
  properly accounted for in our $\chi^2$ statistical analysis described
  above, and are used in the accurate fitting procedure aimed at 
  limiting the model parameter space in what follows.

\subsection{Parameter constraints}
\label{subsec:fit}

First, both selected sets, Zhao and Chuang, are
combined with BAO measurements at low
redshifts (''lowz'' in what follows)
which include MGS and 6dFGS probes, since they are unique in this
redshift range. These probes are 
consistent by itself with the $\Lambda$CDM
cosmology~\cite{Ade:2015xua}~\footnote{In fact, the MGS probe at
  $z=0.15$ gives somewhat higher values of $D_V/r_s$ in comparison
  with the $\Lambda$CDM prediction whereas the DDM decreases it~\cite{Berezhiani:2015yta}, which means that the MGS likelihood would disfavor the DDM model in comparison with the
  $\Lambda$CDM cosmology. This justifies the composition Zhao+lowz as
  the most restrictive pattern to DDM.}. The data sets constructed in
this way are labelled as Zhao+lowz and Chuang+lowz in the analysis below.

Second, we combine Bourboux~\cite{Bourboux:2017cbm} and
Bautista~\cite{Bautista:2017zgn} sets into one sample with the Base
data set to see how much DDM may help to alleviate prominent BAO
Ly-$\alpha$ anomaly. {For this we exploit the $\chi^2$ surface for
  combined measurements available on the
  website (\href{https://github.com/igmhub/picca/tree/master/data}{https://github.com/igmhub/picca/tree/master/data})
  where the Bautista data were extrapolated to the Bourboux's
  effective redshift $z_{\rm eff}=2.4$ using the fiducial cosmology.}

Since DDM is not able to reconcile the
constraints from Ly-$\alpha$ forest absorption with BAO measurements at
middle redshifts completely~\footnote{To resolve the Ly-$\alpha$
  anomaly present in the combined cross- and autocorrelation BAO
  measurements one would need $F=0.1-0.2$ according
  to~\cite{Berezhiani:2015yta}, but such large values are disfavored
  by the Planck likelihood due to strong lensing priors as explained
  in~\cite{Chudaykin:2016yfk}, and, as we already have seen, would
  also put DDM in tension with BAO/RSD probes at middle redshifts.},
we do not mix these inconsistent measurements in one set.
The final choice of the data sets used in our analysis is summarized in
Table\,\ref{tab:sets}. 
\begin{table}[!htb]
\begin{center}
\begin{tabular}{|c| l |}
\hline 
Tag & ~~~~~~~~  ~~~~~~~~ {Data set} \\
\hline 
Zhao+lowz & ~Base+Zhao\cite{Zhao:2016das}+6dFGS\cite{Beutler:2011hx}+MGS\cite{Ross:2014mgs}\\
Chuang+lowz & ~Base+Chuang\cite{Chuang:2016uuz}+6dFGS\cite{Beutler:2011hx}+MGS\cite{Ross:2014mgs}\\
Ly$\alpha$ &  ~Base+Bourboux\cite{Bourboux:2017cbm}+Bautista\cite{Bautista:2017zgn}~\\
\hline 
\end{tabular}
\end{center}
\caption{
Data sets used in our analysis and their tags.
\label{tab:sets}
}
\end{table}

\begin{figure}[!htb]
\includegraphics[keepaspectratio,width=0.9\columnwidth]{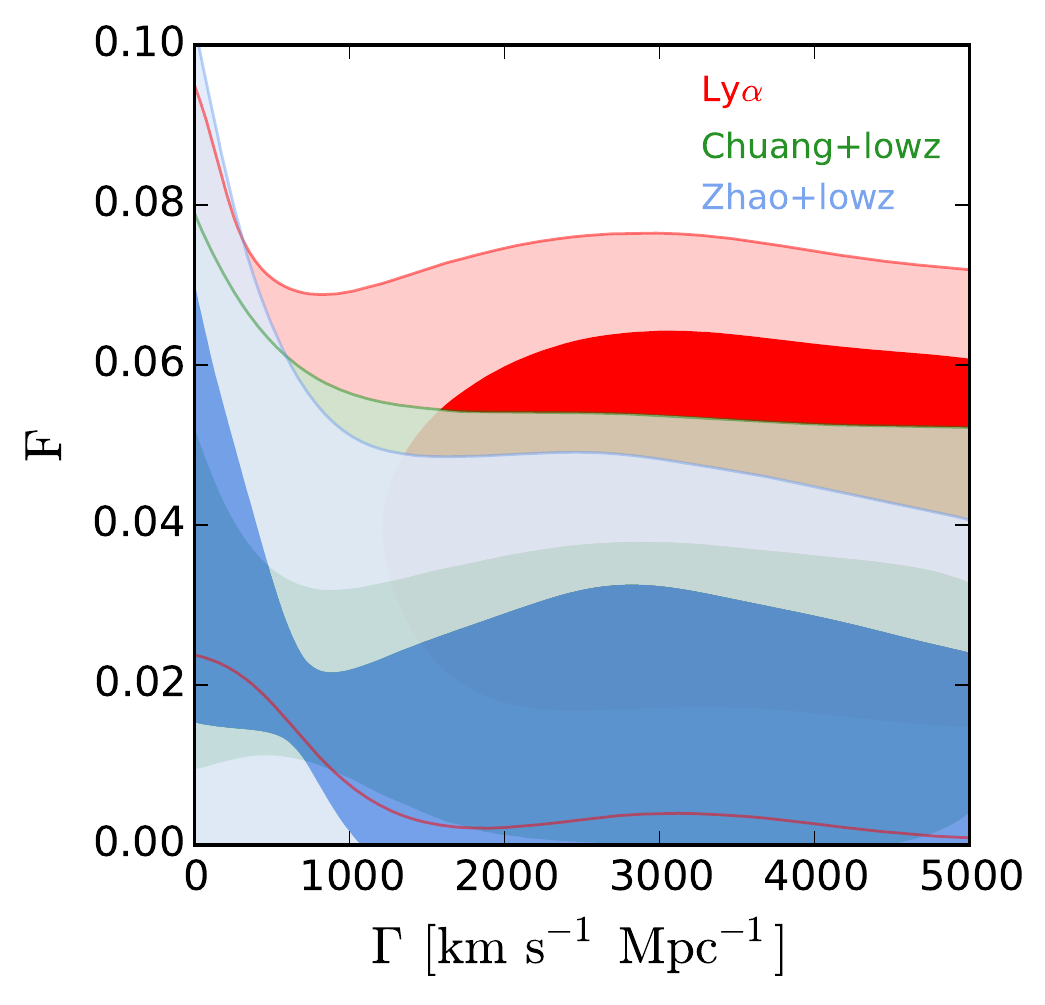}
\caption{\label{fig:F-Gamma} 
Posterior distributions ($1\,\sigma$ and $2\,\sigma$ contours) of parameters $F$,  $\Gamma$ in DDM model. Tags are described in Table~\ref{tab:sets}.}
\end{figure}

\begin{figure}[!htb]
\includegraphics[keepaspectratio,width=0.8\columnwidth]{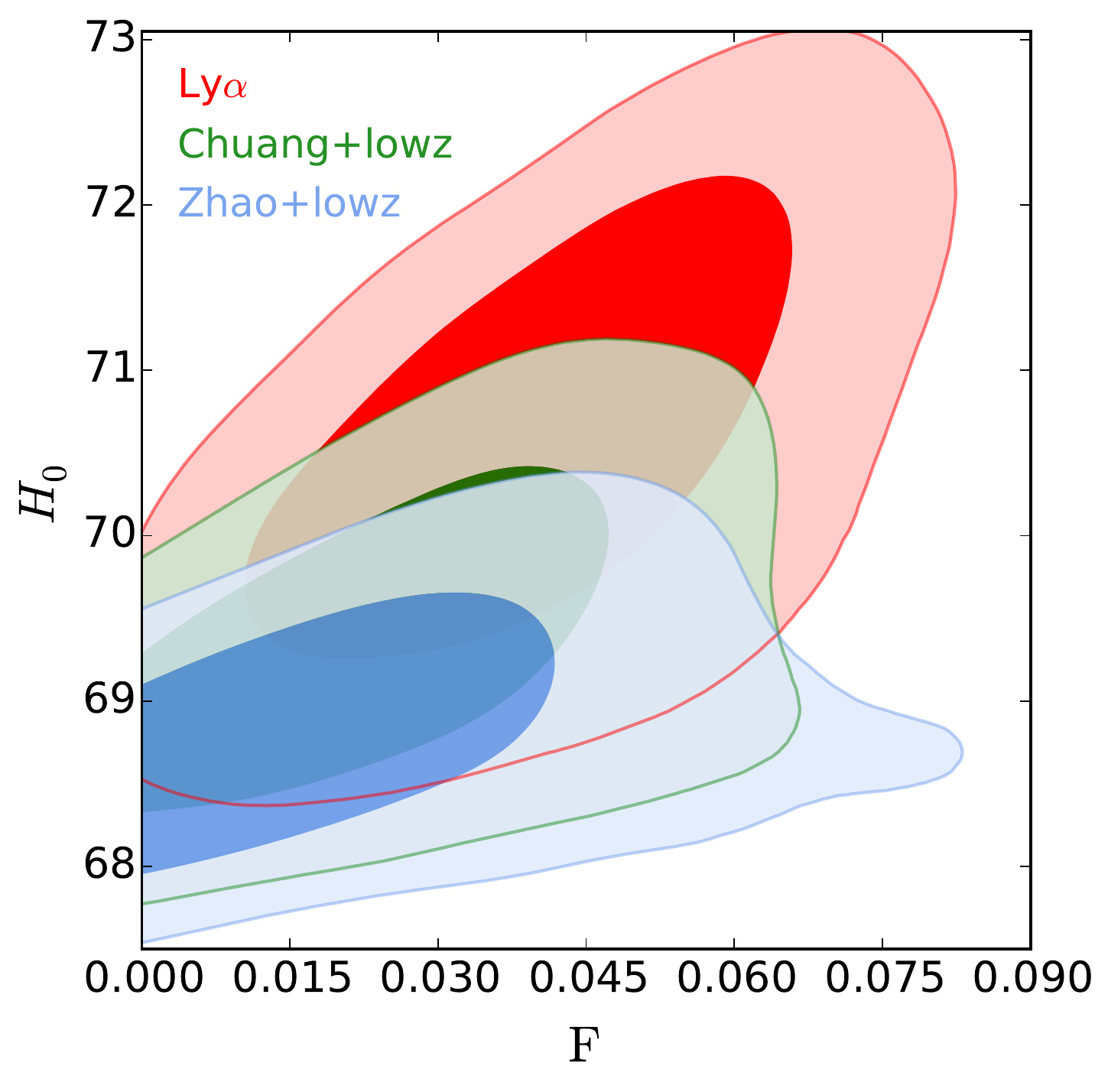}
\caption{\label{fig:H-F} Same as Fig.~\ref{fig:F-Gamma} but for $H_{0}$ and $F$.}
\end{figure}

\begin{figure}[!htb]
\includegraphics[keepaspectratio,width=0.9\columnwidth]{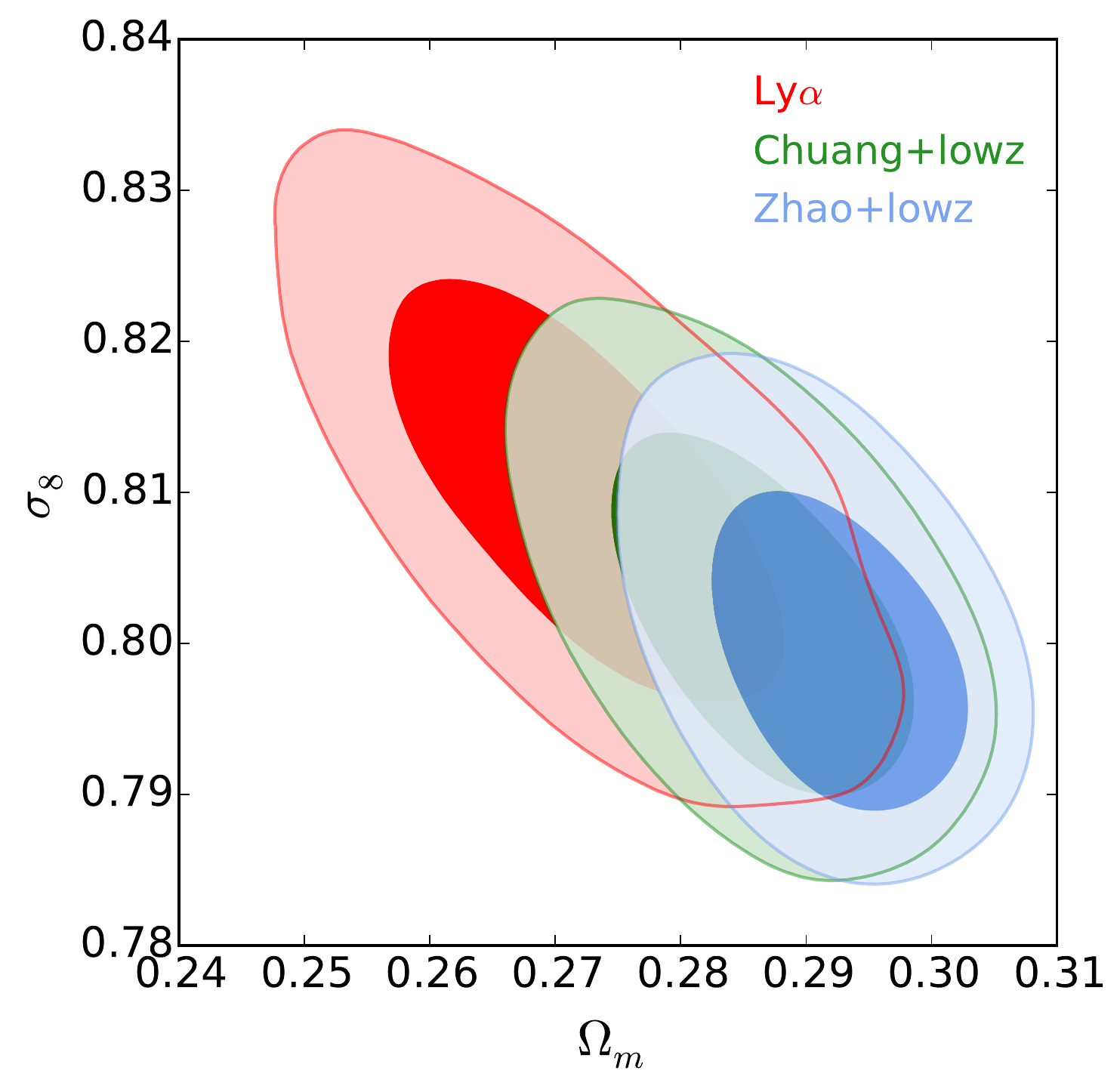}
\caption{\label{fig:sigma8-omegam} 
Same as Fig.~\ref{fig:F-Gamma} but for $\sigma_8$ and $\Omega_m$.} 
\end{figure}

Corresponding constraints in various 2-parameter subspaces are shown
in Fig.~\ref{fig:F-Gamma}, \ref{fig:H-F}, \ref{fig:sigma8-omegam}.

    The probability contours in $\Gamma$, F subspace, Fig.~\ref{fig:F-Gamma}, are almost horizontal at large and intermediate
  values of $\Gamma$, illustrating the simple fact that the
  cosmological data under consideration reflect cosmological processes
  either at very high (CMB) or low (BAO, Hubble measurements, cluster
  counts) redshifts, and therefore possible fine details of the
  Universe evolution at intermediate redshifts are not important.  One may
  also note, that at $1\,\sigma$-level the Ly-$\alpha$ measurements
  actually favor $\Gamma\gg H_0$. Indeed, DDM should decay well
  before the corresponding Ly-$\alpha$ measurements at $z=2.4$ to
  better address BAO signals at high redshifts. One would further
  argue from Fig.~\ref{fig:F-Gamma}, that at $\Gamma\gtrsim H_0$  
  the Zhao+lowz data set exhibits 
  the most pronounced vertical tail, because
  this data set is the most restrictive BAO probes of DDM fraction at
  $z=0.1-0.75$, and hence prefers long-lived DDM. 
  However, our technique is not capable of fully
  resolving such small values of $\Gamma$ and describes quantitatively only the
  short-lived DDM cosmology. In particular, the cosmological data sets 
  related to the nonlinear structures in the late Universe (galaxies
  and galaxy clusters) are obtained from the observations assuming the
  standard $\Lambda$CDM evolution. In our case, a part of DM inside the
  compact structures can decay, which changes their masses and
  gravitational potentials with respect to the standard picture. 
  Anyway, the best-fit values in our study always
  happen at large and intermediate decay rates, and these effects
  are irrelevant.

Presented results confirm anticipated
hierarchy between Zhao+lowz and Chuang+lowz data sets
which impose the
following constraints $F<0.04\,(2\sigma)$ and $F<0.05\,(2\sigma)$,
respectively, while Ly-$\alpha$ measurements lead to
the loosest constraints on the DDM parameter $F<0.07\,(2\sigma)$.
In addition, our constraints imposed by Zhao+lowz and Chuang+lowz data sets are in good agreement with constraints obtained in the short-lived regime in Ref.~\cite{Poulin:2016nat},  whereas Ly$\alpha$ dictates a higher fraction of DDM for the reason explained above.

\begin{figure}[!t]
\includegraphics[keepaspectratio,width=0.9\columnwidth]{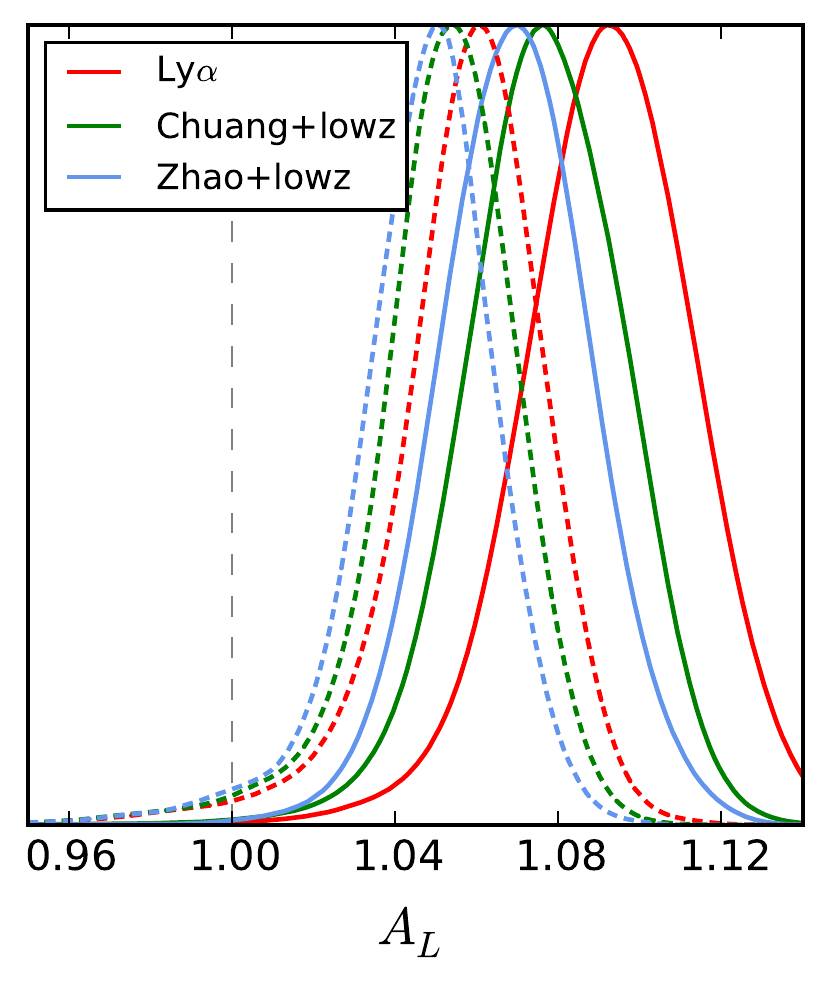}
\caption{\label{fig:AL} 
Marginalized posterior distribution for ${\rm A_L}$ in different data sets described in Table~\ref{tab:sets}. Dashed (solid) lines correspond to constraints obtained in the $\Lambda$CDM  (DDM) model.}
\end{figure}

Since lensing measured by Planck is in conflict with the $\Lambda$CDM
prediction~\cite{Ade:2015xua}, it makes sense to vary a parameter
${\rm A_L}$, which scales the $C_{l}^{\phi\phi}$ power spectrum at
each point in the parameter space. Resulting probability densities for the
lensing power amplitude in $\Lambda$CDM and DDM models using different
data sets listed in Table~\ref{tab:prob} are presented in
Fig.~\ref{fig:AL}. The allowable amount of DDM in this case reaches
the values $F=0.05\pm0.02\,(1\sigma)$, $F=0.06\pm0.02\,(1\sigma)$,
$F=0.07\pm0.02\,(1\sigma)$ for Zhao+lowz, Chuang+lowz
and Ly$\alpha$ data sets, respectively. We emphasize that the nonzero value
of $F$ is preferable now.

Actually, higher values of ${\rm A_L}$ imply a more pronounced
  lensing effect on structures in the late Universe, so a larger
  fraction of DM can decay without conflict with observed CMBR lensing. Indeed, we observed in
  Ref.\,\cite{Chudaykin:2016yfk} that the DDM framework corresponds to a
  weaker lensing power with respect to $\Lambda$CDM. Therefore a more powerful
  lensing in the case of  ${\rm A_L}> 1$ helps to reconcile the tension
  between low- and high-redshift measurements in a more efficient way
 allowing larger fraction $F$.

To understand which model ($\Lambda$CDM or DDM) describes the stack of
current cosmological data better we consider the differences of
logarithmic likelihoods $\log L$ calculated for these two models
within the same data set. The quantity $2\cdot\Delta\log L$ defined in
this way is distributed as $\chi^2$ with $n$ degrees of
freedom equal to the difference in fitting parameters in the models
under consideration. The DDM model has two extra parameters, 
$F$ and $\Gamma$, 
which leads to $n=2$ in our case. Resulting
improvements of the DDM pattern over the $\Lambda$CDM model are listed
in Table~\ref{tab:prob}.

\begin{table}
\begin{center}
\begin{tabular}{|c|c|c|c|}
\hline 
Data set  on top of Base & ~$\Delta \chi^2$~ & ~p-value~ & Improvement\\
\hline 
Zhao+lowz & 0.24 & 0.89 & $0.14\sigma$ \\
Chuang+lowz & 1.88  & 0.39 & $0.86\sigma$ \\
Ly$\alpha$ & 3.94 & 0.14 & $1.48\sigma$ \\
\hline
Zhao+lowz+${\rm A_L}$ & 4.62 & 0.10 & $1.65\sigma$ \\
Chuang+lowz+${\rm A_L}$ & 5.18  & 0.08 & $1.78\sigma$ \\
Ly$\alpha$+${\rm A_L}$ & 13.78 & 0.001 & $3.26\sigma$ \\
\hline 
\end{tabular}
\end{center}
\caption{ Improvement of DDM over $\Lambda$CDM in the three data sets
  considered taking into account 2 extra degrees of freedom in DDM.
\label{tab:prob}
}
\end{table}

\subsection{Discussion}
\label{subsec:discission}

\begin{figure}[!t]
\includegraphics[keepaspectratio,width=0.88\columnwidth]{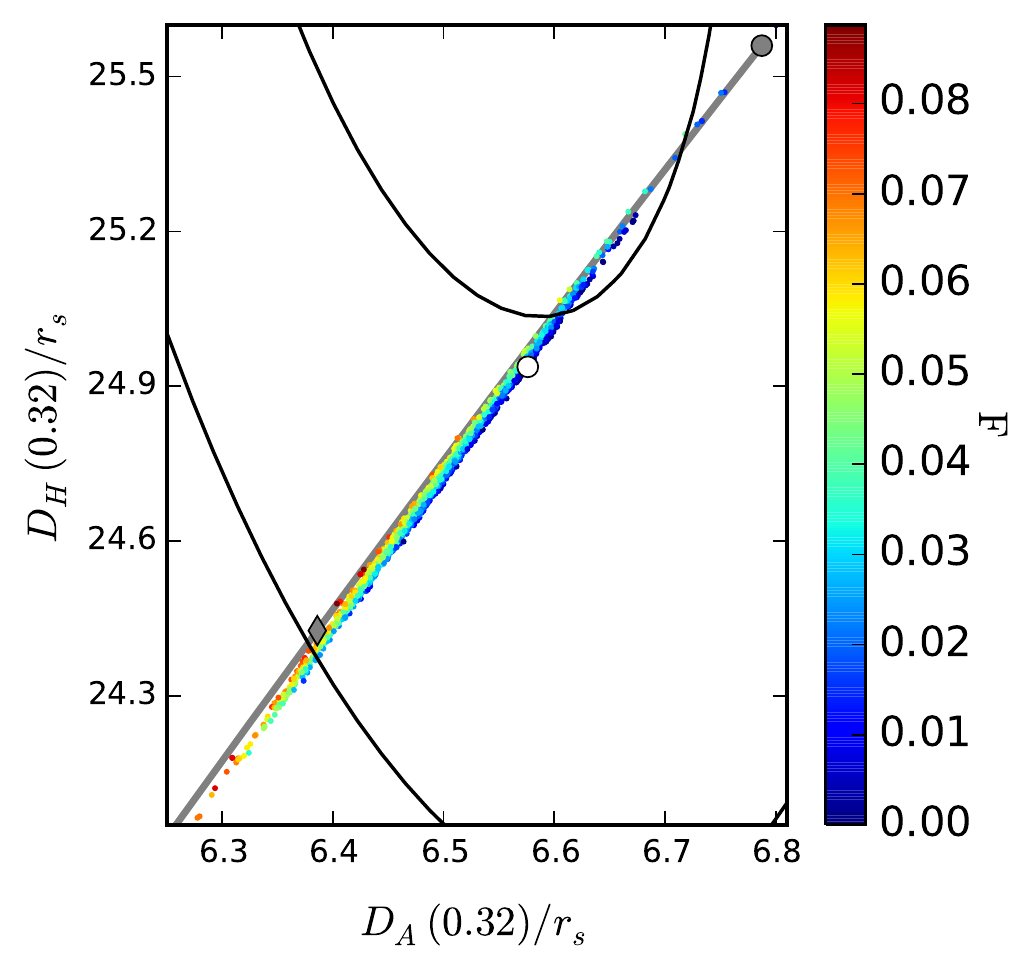}
\caption{\label{fig:LOWZ} Constraints on $D_A/r_s$ and $D_H/r_s$ at
  $z=0.32$ imposed by the Base data set color coded by the value of
  $F$. Black contours show $1\,\sigma$ and $2\,\sigma$ confident
  regions of the Chuang likelihood from the LOWZ galaxy catalogue in
  the redshift range $0.15<z<0.43$. White dot indicates best fit to
  the Base data set in pure $\Lambda$CDM model, while colored
  area around it extends to corresponding 2$\sigma$ confidence
  region. Grey line shows $D_A-D_H$ behavior with the growth of $F$
  in the DDM model assuming $\Gamma=2000$\,km/s/Mpc and all other
  parameters fixed to $\Lambda$CDM-Planck best-fit as in
  Ref.~\cite{Berezhiani:2015yta}. Grey dot and rhombus on this line
  correspond to $F=0$ and $F=0.1$.}
  \vspace{0.2cm}
\end{figure}

\begin{figure}[!t]
\includegraphics[keepaspectratio,width=0.88\columnwidth]{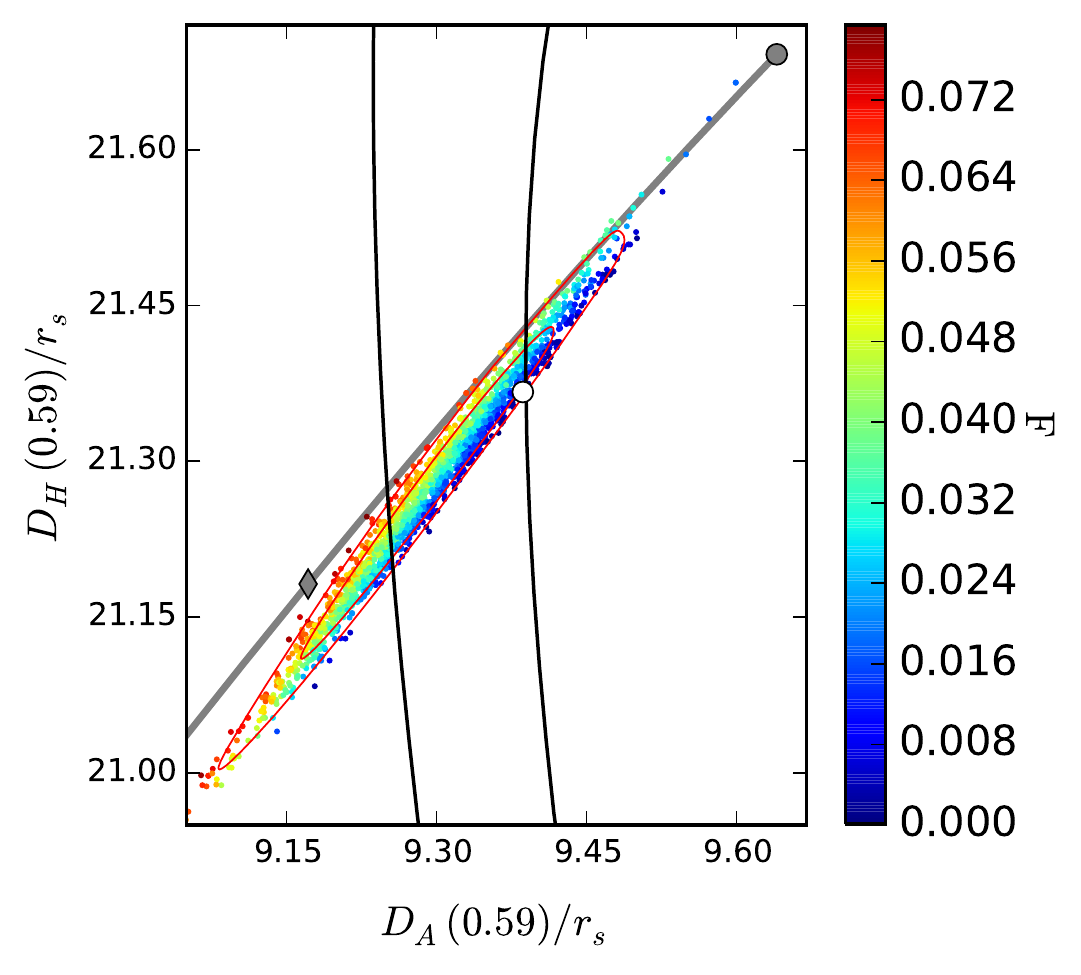}
\caption{\label{fig:CMASS} Same as Fig.~\ref{fig:LOWZ}, but for
  $z=0.59$, and black contours displaying now $1\,\sigma$ and
  $2\,\sigma$ confident regions of the Chuang likelihood from the
  CMASS galaxy catalogue in the redshift range $0.43<z<0.75$. Red
  contours depict $1\,\sigma$ and $2\,\sigma$ confidence regions
  imposed by the Base data set.}
  \vspace{0.96cm}
\end{figure}

\begin{figure}[!t]
\includegraphics[keepaspectratio,width=0.88\columnwidth]{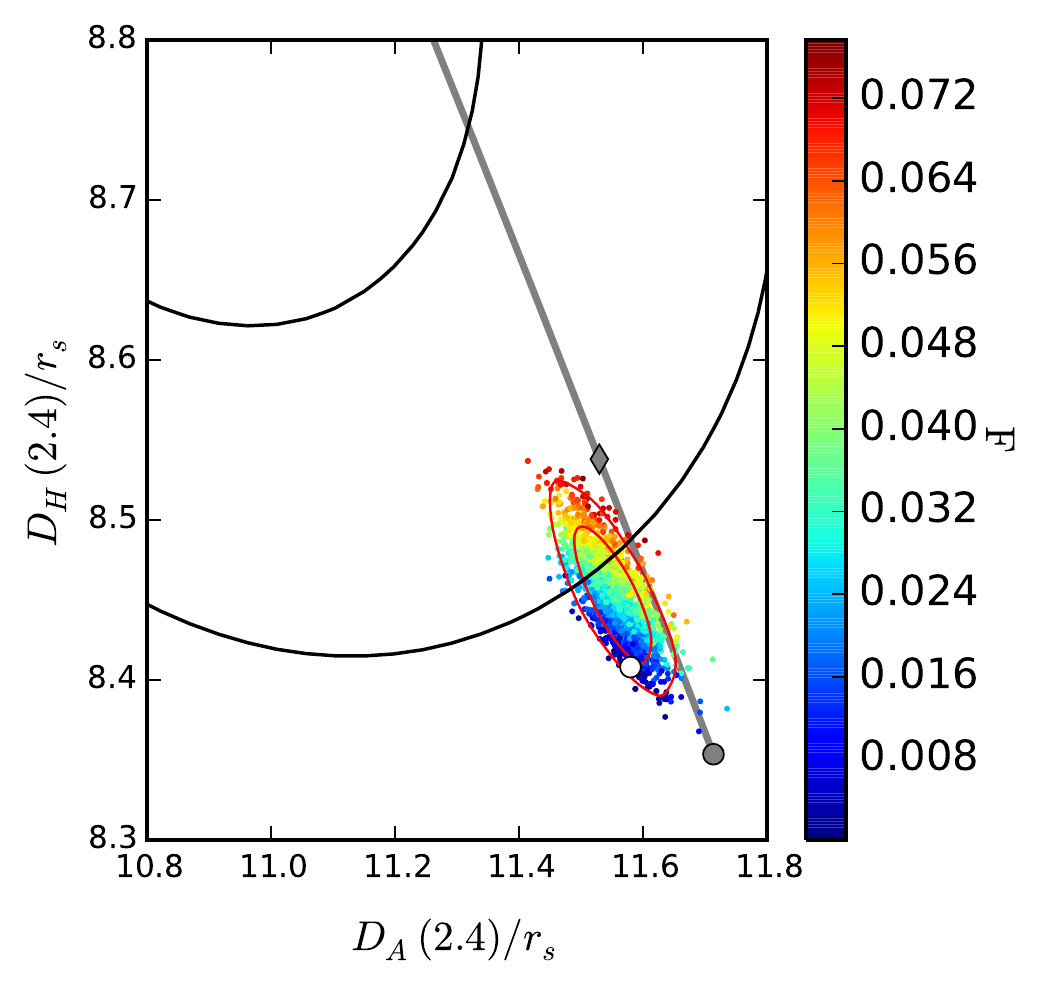}
\caption{Same as Fig.~\ref{fig:LOWZ} but for $z=2.4$, but black contours
  display $1\,\sigma$ and $2\,\sigma$ confident regions of the
  combined likelihood of the Bourboux and
  Bautista Ly-$\alpha$ analyses.}
\label{fig:Lya} 
\end{figure}

To highlight the main results obtained in the paper, let us take a closer
look at the 2d likelihoods in $D_A, D_H$ parameter space. These
likelihoods actually form the basis for the parameter constraints
obtained in Sec.\,\ref{subsec:fit}.  To restrict the number of figures,
we show only Chuang and Ly-$\alpha$ likelihoods: moreover, for the
former, we take likelihoods obtained for $0.15<z<0.43$ and
$0.43<z<0.75$ sample volumes, with mean redshifts 0.32 and 0.59, respectively~\footnote{Parameter constraints of Sec.\,\ref{subsec:fit}
  are obtained with galaxy sample $0.15<z<0.75$ sliced into four
  redshift bins, but for the illustration purposes of the present
  section the two bin splitting, which is also proved by
  Ref.\,\cite{Chuang:2016uuz}, is more appropriate.}.  Those BAO
likelihoods at 1 and 2 $\sigma$ levels are shown in
Figs.\,\ref{fig:LOWZ}-\ref{fig:Lya} by solid black
curves. Constraints on DDM cosmology under the Base data set are
shown by colored areas which extend to their respective 2$\sigma$
confidence levels.  In other words, these colored regions show how
results of Ref.~\cite{Chudaykin:2016yfk} look like on the BAO plane,
with white dot indicating the best fit to the Base data set in
pure $\Lambda$CDM model.

In addition, grey line in Figs.~\ref{fig:LOWZ} - \ref{fig:Lya}
illustrates results of Ref.~\cite{Berezhiani:2015yta}. The grey dot at the
end of the line indicates the best fit of the $\Lambda$CDM model to the
Planck data only. Parameter $F$ of DDM increases from $F = 0$ away
from this point reaching reference value $F=0.1$ at the grey
rhombus. In accordance with anticipation of
Ref.~\cite{Berezhiani:2015yta} the grey line passes near the white dot: 
DDM would resolve the tension between low- and high-z cosmological
data keeping intact the Planck best-fit values to CMB if
lensing of CMB anisotropies could have been neglected. On the other
hand, in $\Lambda$CDM at the white dot, the conflicting low- and high-z data
are ``reconciled'' at the expense of the CMB fit, which deteriorates
somewhat here. But since DDM is worse at the description of lensing, the
overall improvement of DDM over $\Lambda$CDM is not very significant.

Now back to BAO. As one can see from
Figs.~\ref{fig:LOWZ},~\ref{fig:CMASS}, higher values of $F$ are in
conflict with priors provided by the Chuang analysis. This reveals
that not only the lensing of CMB anisotropies observed in the Planck
data severely restricts the DDM model, see~\cite{Chudaykin:2016yfk},
but the middle-redshift measurements of the BOSS galaxy sample are
discordant with the DDM cosmology by themselves. On the contrary,
the combined likelihood of Ly-$\alpha$ forest actually favors higher
values of $F$ according to Fig.~\ref{fig:Lya}, but a corresponding
effect is limited by the Planck lensing priors as explained
in~\cite{Chudaykin:2016yfk}. Still, improvement of DDM over
$\Lambda$CDM is most significant with this data set, see
Table~\ref{tab:prob}.

\section{Conclusions}
\label{sec:conclusions}

To summarize, we have studied the sensitivity of BAO-related
cosmological measurements to the presence of decaying component in the
dark matter sector. We found that all BAO and RSD probes based on the
BOSS DR 12 galaxy sample alone favor the standard $\Lambda$CDM
cosmology. However, the DDM model remains preferable under various
high-z and low-z measurements at the level of
0.1-0.9\,$\sigma$. Employing the BAO information at high redshifts from
cross- and autocorrelations of the Ly-$\alpha$ forest instead we got
even more pronounced 1.5\,$\sigma$ improvement for the DDM model
over the $\Lambda$CDM one.

Since the lensing conflict within the Planck data is not resolved yet
one may consider lensing amplitude ${\rm A_L}$ as a free parameter.
In this case, a nonzero value of $F$ becomes preferable and the DDM
scenario improves the goodness-of-fit by 1.7-3.3\,$\sigma$ in
comparison with the concordance $\Lambda$CDM model. The improvement
depends on a particular choice of the BAO/RSD measurements included in
the analysis.

Fixing the origin of deviations between best-fit cosmological values
of various BAO/RSD cosmological probes would strengthen our
conclusions either in favor or against the decaying dark matter
component.  Ongoing and future galaxy survey projects like DES,
  EUCLID, LSST will most probably help with this problem, while
  planning 21\,cm intensity mapping surveys might even probe the 
relatively short and intermediate lifetimes $\Gamma^{-1}$ of the decaying
  component, which 
  escape any grasps of the present cosmological analysis. The 
  available cosmological data place an upper limit on the DDM fraction
  $F$ of 4-7\% at 95\% CL, while some probes hint the presence of the
  same DDM fraction at the same CL.

\acknowledgements{
We thank J. Bautista, H. du Mas des Bourboux and Y. Wang who shared the data with us making a valuable contribution to this work. We are also grateful to Sh. Alam, F. Beutler, J. Grieb, M. Pellejero-Ibanez, A. Ross and S. Satpathy for discussions and  information about analysis procedure. All computations in the work were made with the MVS-10P supercomputer of the Joint Supercomputer Center of the Russian Academy of Sciences (JSCC RAS).
The work has been supported by the RSF grant 14-22-00161. }

\bibliography{refs} 

\begin{thebibliography}{31}%
\makeatletter
\providecommand \@ifxundefined [1]{%
 \@ifx{#1\undefined}
}%
\providecommand \@ifnum [1]{%
 \ifnum #1\expandafter \@firstoftwo
 \else \expandafter \@secondoftwo
 \fi
}%
\providecommand \@ifx [1]{%
 \ifx #1\expandafter \@firstoftwo
 \else \expandafter \@secondoftwo
 \fi
}%
\providecommand \natexlab [1]{#1}%
\providecommand \enquote  [1]{``#1''}%
\providecommand \bibnamefont  [1]{#1}%
\providecommand \bibfnamefont [1]{#1}%
\providecommand \citenamefont [1]{#1}%
\providecommand \href@noop [0]{\@secondoftwo}%
\providecommand \href [0]{\begingroup \@sanitize@url \@href}%
\providecommand \@href[1]{\@@startlink{#1}\@@href}%
\providecommand \@@href[1]{\endgroup#1\@@endlink}%
\providecommand \@sanitize@url [0]{\catcode `\\12\catcode `\$12\catcode
  `\&12\catcode `\#12\catcode `\^12\catcode `\_12\catcode `\%12\relax}%
\providecommand \@@startlink[1]{}%
\providecommand \@@endlink[0]{}%
\providecommand \url  [0]{\begingroup\@sanitize@url \@url }%
\providecommand \@url [1]{\endgroup\@href {#1}{\urlprefix }}%
\providecommand \urlprefix  [0]{URL }%
\providecommand \Eprint [0]{\href }%
\providecommand \doibase [0]{http://dx.doi.org/}%
\providecommand \selectlanguage [0]{\@gobble}%
\providecommand \bibinfo  [0]{\@secondoftwo}%
\providecommand \bibfield  [0]{\@secondoftwo}%
\providecommand \translation [1]{[#1]}%
\providecommand \BibitemOpen [0]{}%
\providecommand \bibitemStop [0]{}%
\providecommand \bibitemNoStop [0]{.\EOS\space}%
\providecommand \EOS [0]{\spacefactor3000\relax}%
\providecommand \BibitemShut  [1]{\csname bibitem#1\endcsname}%
\let\auto@bib@innerbib\@empty
\bibitem [{\citenamefont {Berezhiani}\ \emph {et~al.}(2015)\citenamefont
  {Berezhiani}, \citenamefont {Dolgov},\ and\ \citenamefont
  {Tkachev}}]{Berezhiani:2015yta}%
  \BibitemOpen
  \bibfield  {author} {\bibinfo {author} {\bibfnamefont {Z.}~\bibnamefont
  {Berezhiani}}, \bibinfo {author} {\bibfnamefont {A.~D.}\ \bibnamefont
  {Dolgov}}, \ and\ \bibinfo {author} {\bibfnamefont {I.~I.}\ \bibnamefont
  {Tkachev}},\ }\href {\doibase 10.1103/PhysRevD.92.061303} {\bibfield
  {journal} {\bibinfo  {journal} {Phys. Rev.}\ }\textbf {\bibinfo {volume}
  {D92}},\ \bibinfo {pages} {061303} (\bibinfo {year} {2015})},\ \Eprint
  {http://arxiv.org/abs/1505.03644} {arXiv:1505.03644 [astro-ph.CO]}
  \BibitemShut {NoStop}%
\bibitem [{\citenamefont {Chudaykin}\ \emph {et~al.}(2016)\citenamefont
  {Chudaykin}, \citenamefont {Gorbunov},\ and\ \citenamefont
  {Tkachev}}]{Chudaykin:2016yfk}%
  \BibitemOpen
  \bibfield  {author} {\bibinfo {author} {\bibfnamefont {A.}~\bibnamefont
  {Chudaykin}}, \bibinfo {author} {\bibfnamefont {D.}~\bibnamefont {Gorbunov}},
  \ and\ \bibinfo {author} {\bibfnamefont {I.}~\bibnamefont {Tkachev}},\ }\href
  {\doibase 10.1103/PhysRevD.94.023528} {\bibfield  {journal} {\bibinfo
  {journal} {Phys. Rev.}\ }\textbf {\bibinfo {volume} {D94}},\ \bibinfo {pages}
  {023528} (\bibinfo {year} {2016})},\ \Eprint
  {http://arxiv.org/abs/1602.08121} {arXiv:1602.08121 [astro-ph.CO]}
  \BibitemShut {NoStop}%
\bibitem [{\citenamefont {Flores}\ \emph {et~al.}(1986)\citenamefont {Flores},
  \citenamefont {Blumenthal}, \citenamefont {Dekel},\ and\ \citenamefont
  {Primack}}]{Flores:1986jn}%
  \BibitemOpen
  \bibfield  {author} {\bibinfo {author} {\bibfnamefont {R.}~\bibnamefont
  {Flores}}, \bibinfo {author} {\bibfnamefont {G.~R.}\ \bibnamefont
  {Blumenthal}}, \bibinfo {author} {\bibfnamefont {A.}~\bibnamefont {Dekel}}, \
  and\ \bibinfo {author} {\bibfnamefont {J.~R.}\ \bibnamefont {Primack}},\
  }\href {\doibase 10.1038/323781a0} {\bibfield  {journal} {\bibinfo  {journal}
  {Nature}\ }\textbf {\bibinfo {volume} {323}},\ \bibinfo {pages} {781}
  (\bibinfo {year} {1986})}\BibitemShut {NoStop}%
\bibitem [{\citenamefont {Doroshkevich}\ \emph {et~al.}(1989)\citenamefont
  {Doroshkevich}, \citenamefont {Khlopov},\ and\ \citenamefont
  {Klypin}}]{Doroshkevich:1989bf}%
  \BibitemOpen
  \bibfield  {author} {\bibinfo {author} {\bibfnamefont {A.~G.}\ \bibnamefont
  {Doroshkevich}}, \bibinfo {author} {\bibfnamefont {M.}~\bibnamefont
  {Khlopov}}, \ and\ \bibinfo {author} {\bibfnamefont {A.~A.}\ \bibnamefont
  {Klypin}},\ }\href@noop {} {\bibfield  {journal} {\bibinfo  {journal} {Mon.
  Not. Roy. Astron. Soc.}\ }\textbf {\bibinfo {volume} {239}},\ \bibinfo
  {pages} {923} (\bibinfo {year} {1989})}\BibitemShut {NoStop}%
\bibitem [{\citenamefont {Ade}\ \emph {et~al.}(2014)\citenamefont {Ade} \emph
  {et~al.}}]{Ade:2013zuv}%
  \BibitemOpen
  \bibfield  {author} {\bibinfo {author} {\bibfnamefont {P.~A.~R.}\
  \bibnamefont {Ade}} \emph {et~al.} (\bibinfo {collaboration} {Planck}),\
  }\href {\doibase 10.1051/0004-6361/201321591} {\bibfield  {journal} {\bibinfo
   {journal} {Astron. Astrophys.}\ }\textbf {\bibinfo {volume} {571}},\
  \bibinfo {pages} {A16} (\bibinfo {year} {2014})},\ \Eprint
  {http://arxiv.org/abs/1303.5076} {arXiv:1303.5076 [astro-ph.CO]} \BibitemShut
  {NoStop}%
\bibitem [{\citenamefont {Ade}\ \emph {et~al.}(2015{\natexlab{a}})\citenamefont
  {Ade} \emph {et~al.}}]{Ade:2015xua}%
  \BibitemOpen
  \bibfield  {author} {\bibinfo {author} {\bibfnamefont {P.~A.~R.}\
  \bibnamefont {Ade}} \emph {et~al.} (\bibinfo {collaboration} {Planck}),\
  }\href@noop {} {\  (\bibinfo {year} {2015}{\natexlab{a}})},\ \Eprint
  {http://arxiv.org/abs/1502.01589} {arXiv:1502.01589 [astro-ph.CO]}
  \BibitemShut {NoStop}%
\bibitem [{\citenamefont {Riess}\ \emph {et~al.}(2011)\citenamefont {Riess},
  \citenamefont {Macri}, \citenamefont {Casertano}, \citenamefont {Lampeitl},
  \citenamefont {Ferguson}, \citenamefont {Filippenko}, \citenamefont {Jha},
  \citenamefont {Li},\ and\ \citenamefont {Chornock}}]{Riess:2011yx}%
  \BibitemOpen
  \bibfield  {author} {\bibinfo {author} {\bibfnamefont {A.~G.}\ \bibnamefont
  {Riess}}, \bibinfo {author} {\bibfnamefont {L.}~\bibnamefont {Macri}},
  \bibinfo {author} {\bibfnamefont {S.}~\bibnamefont {Casertano}}, \bibinfo
  {author} {\bibfnamefont {H.}~\bibnamefont {Lampeitl}}, \bibinfo {author}
  {\bibfnamefont {H.~C.}\ \bibnamefont {Ferguson}}, \bibinfo {author}
  {\bibfnamefont {A.~V.}\ \bibnamefont {Filippenko}}, \bibinfo {author}
  {\bibfnamefont {S.~W.}\ \bibnamefont {Jha}}, \bibinfo {author} {\bibfnamefont
  {W.}~\bibnamefont {Li}}, \ and\ \bibinfo {author} {\bibfnamefont
  {R.}~\bibnamefont {Chornock}},\ }\href {\doibase 10.1088/0004-637X/732/2/129,
  10.1088/0004-637X/730/2/119} {\bibfield  {journal} {\bibinfo  {journal}
  {Astrophys. J.}\ }\textbf {\bibinfo {volume} {730}},\ \bibinfo {pages} {119}
  (\bibinfo {year} {2011})},\ \bibinfo {note} {[Erratum: Astrophys.
  J.732,129(2011)]},\ \Eprint {http://arxiv.org/abs/1103.2976} {arXiv:1103.2976
  [astro-ph.CO]} \BibitemShut {NoStop}%
\bibitem [{\citenamefont {Ade}\ \emph {et~al.}(2015{\natexlab{b}})\citenamefont
  {Ade} \emph {et~al.}}]{Ade:2015fva}%
  \BibitemOpen
  \bibfield  {author} {\bibinfo {author} {\bibfnamefont {P.~A.~R.}\
  \bibnamefont {Ade}} \emph {et~al.} (\bibinfo {collaboration} {Planck}),\
  }\href@noop {} {\  (\bibinfo {year} {2015}{\natexlab{b}})},\ \Eprint
  {http://arxiv.org/abs/1502.01597} {arXiv:1502.01597 [astro-ph.CO]}
  \BibitemShut {NoStop}%
\bibitem [{\citenamefont {du~Mas~des Bourboux}\ \emph
  {et~al.}(2017)\citenamefont {du~Mas~des Bourboux} \emph
  {et~al.}}]{Bourboux:2017cbm}%
  \BibitemOpen
  \bibfield  {author} {\bibinfo {author} {\bibfnamefont {H.}~\bibnamefont
  {du~Mas~des Bourboux}} \emph {et~al.},\ }\href@noop {} {\  (\bibinfo {year}
  {2017})},\ \Eprint {http://arxiv.org/abs/1708.02225} {arXiv:1708.02225
  [astro-ph.CO]} \BibitemShut {NoStop}%
\bibitem [{\citenamefont {Evslin}(2016)}]{Evslin:2015uwa}%
  \BibitemOpen
  \bibfield  {author} {\bibinfo {author} {\bibfnamefont {J.}~\bibnamefont
  {Evslin}},\ }\href {\doibase 10.1016/j.dark.2016.06.001} {\bibfield
  {journal} {\bibinfo  {journal} {Phys. Dark Univ.}\ }\textbf {\bibinfo
  {volume} {13}},\ \bibinfo {pages} {126} (\bibinfo {year} {2016})},\ \Eprint
  {http://arxiv.org/abs/1510.05630} {arXiv:1510.05630 [astro-ph.CO]}
  \BibitemShut {NoStop}%
\bibitem [{\citenamefont {Evslin}(2017)}]{Evslin:2016gre}%
  \BibitemOpen
  \bibfield  {author} {\bibinfo {author} {\bibfnamefont {J.}~\bibnamefont
  {Evslin}},\ }\href {\doibase 10.1088/1475-7516/2017/04/024} {\bibfield
  {journal} {\bibinfo  {journal} {JCAP}\ }\textbf {\bibinfo {volume} {1704}},\
  \bibinfo {pages} {024} (\bibinfo {year} {2017})},\ \Eprint
  {http://arxiv.org/abs/1604.02809} {arXiv:1604.02809 [astro-ph.CO]}
  \BibitemShut {NoStop}%
\bibitem [{\citenamefont {Ade}\ \emph {et~al.}(2015{\natexlab{c}})\citenamefont
  {Ade} \emph {et~al.}}]{Ade:2015len}%
  \BibitemOpen
  \bibfield  {author} {\bibinfo {author} {\bibfnamefont {P.~A.~R.}\
  \bibnamefont {Ade}} \emph {et~al.} (\bibinfo {collaboration} {Planck}),\
  }\href@noop {} {\  (\bibinfo {year} {2015}{\natexlab{c}})},\ \Eprint
  {http://arxiv.org/abs/1502.01591} {arXiv:1502.01591 [astro-ph.CO]}
  \BibitemShut {NoStop}%
\bibitem [{\citenamefont {Adam}\ \emph {et~al.}(2016)\citenamefont {Adam} \emph
  {et~al.}}]{Adam:2016hgk}%
  \BibitemOpen
  \bibfield  {author} {\bibinfo {author} {\bibfnamefont {R.}~\bibnamefont
  {Adam}} \emph {et~al.} (\bibinfo {collaboration} {Planck}),\ }\href {\doibase
  10.1051/0004-6361/201628897} {\bibfield  {journal} {\bibinfo  {journal}
  {Astron. Astrophys.}\ }\textbf {\bibinfo {volume} {596}},\ \bibinfo {pages}
  {A108} (\bibinfo {year} {2016})},\ \Eprint {http://arxiv.org/abs/1605.03507}
  {arXiv:1605.03507 [astro-ph.CO]} \BibitemShut {NoStop}%
\bibitem [{\citenamefont {Riess}\ \emph {et~al.}(2016)\citenamefont {Riess}
  \emph {et~al.}}]{Riess:2016jrr}%
  \BibitemOpen
  \bibfield  {author} {\bibinfo {author} {\bibfnamefont {A.~G.}\ \bibnamefont
  {Riess}} \emph {et~al.},\ }\href {\doibase 10.3847/0004-637X/826/1/56}
  {\bibfield  {journal} {\bibinfo  {journal} {Astrophys. J.}\ }\textbf
  {\bibinfo {volume} {826}},\ \bibinfo {pages} {56} (\bibinfo {year} {2016})},\
  \Eprint {http://arxiv.org/abs/1604.01424} {arXiv:1604.01424 [astro-ph.CO]}
  \BibitemShut {NoStop}%
\bibitem [{\citenamefont {Ross}\ \emph
  {et~al.}(2015{\natexlab{a}})\citenamefont {Ross}, \citenamefont {Samushia},
  \citenamefont {Howlett}, \citenamefont {Percival}, \citenamefont {Burden},\
  and\ \citenamefont {Manera}}]{Ross:2014qpa}%
  \BibitemOpen
  \bibfield  {author} {\bibinfo {author} {\bibfnamefont {A.~J.}\ \bibnamefont
  {Ross}}, \bibinfo {author} {\bibfnamefont {L.}~\bibnamefont {Samushia}},
  \bibinfo {author} {\bibfnamefont {C.}~\bibnamefont {Howlett}}, \bibinfo
  {author} {\bibfnamefont {W.~J.}\ \bibnamefont {Percival}}, \bibinfo {author}
  {\bibfnamefont {A.}~\bibnamefont {Burden}}, \ and\ \bibinfo {author}
  {\bibfnamefont {M.}~\bibnamefont {Manera}},\ }\href {\doibase
  10.1093/mnras/stv154} {\bibfield  {journal} {\bibinfo  {journal} {Mon. Not.
  Roy. Astron. Soc.}\ }\textbf {\bibinfo {volume} {449}},\ \bibinfo {pages}
  {835} (\bibinfo {year} {2015}{\natexlab{a}})},\ \Eprint
  {http://arxiv.org/abs/1409.3242} {arXiv:1409.3242 [astro-ph.CO]} \BibitemShut
  {NoStop}%
\bibitem [{\citenamefont {Beutler}\ \emph {et~al.}(2011)\citenamefont
  {Beutler}, \citenamefont {Blake}, \citenamefont {Colless}, \citenamefont
  {Jones}, \citenamefont {Staveley-Smith}, \citenamefont {Campbell},
  \citenamefont {Parker}, \citenamefont {Saunders},\ and\ \citenamefont
  {Watson}}]{Beutler:2011hx}%
  \BibitemOpen
  \bibfield  {author} {\bibinfo {author} {\bibfnamefont {F.}~\bibnamefont
  {Beutler}}, \bibinfo {author} {\bibfnamefont {C.}~\bibnamefont {Blake}},
  \bibinfo {author} {\bibfnamefont {M.}~\bibnamefont {Colless}}, \bibinfo
  {author} {\bibfnamefont {D.~H.}\ \bibnamefont {Jones}}, \bibinfo {author}
  {\bibfnamefont {L.}~\bibnamefont {Staveley-Smith}}, \bibinfo {author}
  {\bibfnamefont {L.}~\bibnamefont {Campbell}}, \bibinfo {author}
  {\bibfnamefont {Q.}~\bibnamefont {Parker}}, \bibinfo {author} {\bibfnamefont
  {W.}~\bibnamefont {Saunders}}, \ and\ \bibinfo {author} {\bibfnamefont
  {F.}~\bibnamefont {Watson}},\ }\href {\doibase
  10.1111/j.1365-2966.2011.19250.x} {\bibfield  {journal} {\bibinfo  {journal}
  {Mon. Not. Roy. Astron. Soc.}\ }\textbf {\bibinfo {volume} {416}},\ \bibinfo
  {pages} {3017} (\bibinfo {year} {2011})},\ \Eprint
  {http://arxiv.org/abs/1106.3366} {arXiv:1106.3366 [astro-ph.CO]} \BibitemShut
  {NoStop}%
\bibitem [{\citenamefont {Cuesta}\ \emph {et~al.}(2016)\citenamefont {Cuesta}
  \emph {et~al.}}]{Cuesta:2015mqa}%
  \BibitemOpen
  \bibfield  {author} {\bibinfo {author} {\bibfnamefont {A.~J.}\ \bibnamefont
  {Cuesta}} \emph {et~al.},\ }\href {\doibase 10.1093/mnras/stw066} {\bibfield
  {journal} {\bibinfo  {journal} {Mon. Not. Roy. Astron. Soc.}\ }\textbf
  {\bibinfo {volume} {457}},\ \bibinfo {pages} {1770} (\bibinfo {year}
  {2016})},\ \Eprint {http://arxiv.org/abs/1509.06371} {arXiv:1509.06371
  [astro-ph.CO]} \BibitemShut {NoStop}%
\bibitem [{\citenamefont {Gil-Marín}\ \emph
  {et~al.}(2016{\natexlab{a}})\citenamefont {Gil-Marín} \emph
  {et~al.}}]{Gil-Marin:2015nqa}%
  \BibitemOpen
  \bibfield  {author} {\bibinfo {author} {\bibfnamefont {H.}~\bibnamefont
  {Gil-Marín}} \emph {et~al.},\ }\href {\doibase 10.1093/mnras/stw1264}
  {\bibfield  {journal} {\bibinfo  {journal} {Mon. Not. Roy. Astron. Soc.}\
  }\textbf {\bibinfo {volume} {460}},\ \bibinfo {pages} {4210} (\bibinfo {year}
  {2016}{\natexlab{a}})},\ \Eprint {http://arxiv.org/abs/1509.06373}
  {arXiv:1509.06373 [astro-ph.CO]} \BibitemShut {NoStop}%
\bibitem [{\citenamefont {Alam}\ \emph {et~al.}(2016)\citenamefont {Alam} \emph
  {et~al.}}]{Alam:2016hwk}%
  \BibitemOpen
  \bibfield  {author} {\bibinfo {author} {\bibfnamefont {S.}~\bibnamefont
  {Alam}} \emph {et~al.} (\bibinfo {collaboration} {BOSS}),\ }\href@noop {}
  {\bibfield  {journal} {\bibinfo  {journal} {Submitted to: Mon. Not. Roy.
  Astron. Soc.}\ } (\bibinfo {year} {2016})},\ \Eprint
  {http://arxiv.org/abs/1607.03155} {arXiv:1607.03155 [astro-ph.CO]}
  \BibitemShut {NoStop}%
\bibitem [{\citenamefont {Zhao}\ \emph {et~al.}(2017)\citenamefont {Zhao} \emph
  {et~al.}}]{Zhao:2016das}%
  \BibitemOpen
  \bibfield  {author} {\bibinfo {author} {\bibfnamefont {G.-B.}\ \bibnamefont
  {Zhao}} \emph {et~al.} (\bibinfo {collaboration} {BOSS}),\ }\href {\doibase
  10.1093/mnras/stw3199} {\bibfield  {journal} {\bibinfo  {journal} {Mon. Not.
  Roy. Astron. Soc.}\ }\textbf {\bibinfo {volume} {466}},\ \bibinfo {pages}
  {762} (\bibinfo {year} {2017})},\ \Eprint {http://arxiv.org/abs/1607.03153}
  {arXiv:1607.03153 [astro-ph.CO]} \BibitemShut {NoStop}%
\bibitem [{\citenamefont {Wang}\ \emph {et~al.}(2016)\citenamefont {Wang} \emph
  {et~al.}}]{Wang:2016wjr}%
  \BibitemOpen
  \bibfield  {author} {\bibinfo {author} {\bibfnamefont {Y.}~\bibnamefont
  {Wang}} \emph {et~al.} (\bibinfo {collaboration} {BOSS}),\ }\href@noop {}
  {\bibfield  {journal} {\bibinfo  {journal} {Submitted to: Mon. Not. Roy.
  Astron. Soc.}\ } (\bibinfo {year} {2016})},\ \bibinfo {note} {[Mon. Not. Roy.
  Astron. Soc.469,3762(2017)]},\ \Eprint {http://arxiv.org/abs/1607.03154}
  {arXiv:1607.03154 [astro-ph.CO]} \BibitemShut {NoStop}%
\bibitem [{\citenamefont {Chuang}\ \emph {et~al.}(2016)\citenamefont {Chuang}
  \emph {et~al.}}]{Chuang:2016uuz}%
  \BibitemOpen
  \bibfield  {author} {\bibinfo {author} {\bibfnamefont {C.-H.}\ \bibnamefont
  {Chuang}} \emph {et~al.} (\bibinfo {collaboration} {BOSS}),\ }\href@noop {}
  {\bibfield  {journal} {\bibinfo  {journal} {Submitted to: Mon. Not. Roy.
  Astron. Soc.}\ } (\bibinfo {year} {2016})},\ \Eprint
  {http://arxiv.org/abs/1607.03151} {arXiv:1607.03151 [astro-ph.CO]}
  \BibitemShut {NoStop}%
\bibitem [{\citenamefont {Gil-Marín}\ \emph
  {et~al.}(2016{\natexlab{b}})\citenamefont {Gil-Marín} \emph
  {et~al.}}]{Gil-Marin:2015sqa}%
  \BibitemOpen
  \bibfield  {author} {\bibinfo {author} {\bibfnamefont {H.}~\bibnamefont
  {Gil-Marín}} \emph {et~al.},\ }\href {\doibase 10.1093/mnras/stw1096}
  {\bibfield  {journal} {\bibinfo  {journal} {Mon. Not. Roy. Astron. Soc.}\
  }\textbf {\bibinfo {volume} {460}},\ \bibinfo {pages} {4188} (\bibinfo {year}
  {2016}{\natexlab{b}})},\ \Eprint {http://arxiv.org/abs/1509.06386}
  {arXiv:1509.06386 [astro-ph.CO]} \BibitemShut {NoStop}%
\bibitem [{\citenamefont {Font-Ribera}\ \emph {et~al.}(2014)\citenamefont
  {Font-Ribera} \emph {et~al.}}]{Font-Ribera:2013wce}%
  \BibitemOpen
  \bibfield  {author} {\bibinfo {author} {\bibfnamefont {A.}~\bibnamefont
  {Font-Ribera}} \emph {et~al.} (\bibinfo {collaboration} {BOSS}),\ }\href
  {\doibase 10.1088/1475-7516/2014/05/027} {\bibfield  {journal} {\bibinfo
  {journal} {JCAP}\ }\textbf {\bibinfo {volume} {1405}},\ \bibinfo {pages}
  {027} (\bibinfo {year} {2014})},\ \Eprint {http://arxiv.org/abs/1311.1767}
  {arXiv:1311.1767 [astro-ph.CO]} \BibitemShut {NoStop}%
\bibitem [{\citenamefont {Delubac}\ \emph {et~al.}(2015)\citenamefont {Delubac}
  \emph {et~al.}}]{Delubac:2014aqe}%
  \BibitemOpen
  \bibfield  {author} {\bibinfo {author} {\bibfnamefont {T.}~\bibnamefont
  {Delubac}} \emph {et~al.} (\bibinfo {collaboration} {BOSS}),\ }\href
  {\doibase 10.1051/0004-6361/201423969} {\bibfield  {journal} {\bibinfo
  {journal} {Astron. Astrophys.}\ }\textbf {\bibinfo {volume} {574}},\ \bibinfo
  {pages} {A59} (\bibinfo {year} {2015})},\ \Eprint
  {http://arxiv.org/abs/1404.1801} {arXiv:1404.1801 [astro-ph.CO]} \BibitemShut
  {NoStop}%
\bibitem [{\citenamefont {Bautista}\ \emph {et~al.}(2017)\citenamefont
  {Bautista} \emph {et~al.}}]{Bautista:2017zgn}%
  \BibitemOpen
  \bibfield  {author} {\bibinfo {author} {\bibfnamefont {J.~E.}\ \bibnamefont
  {Bautista}} \emph {et~al.},\ }\href@noop {} {\  (\bibinfo {year} {2017})},\
  \Eprint {http://arxiv.org/abs/1702.00176} {arXiv:1702.00176 [astro-ph.CO]}
  \BibitemShut {NoStop}%
\bibitem [{\citenamefont {Audren}\ \emph {et~al.}(2013)\citenamefont {Audren},
  \citenamefont {Lesgourgues}, \citenamefont {Benabed},\ and\ \citenamefont
  {Prunet}}]{Audren:2012wb}%
  \BibitemOpen
  \bibfield  {author} {\bibinfo {author} {\bibfnamefont {B.}~\bibnamefont
  {Audren}}, \bibinfo {author} {\bibfnamefont {J.}~\bibnamefont {Lesgourgues}},
  \bibinfo {author} {\bibfnamefont {K.}~\bibnamefont {Benabed}}, \ and\
  \bibinfo {author} {\bibfnamefont {S.}~\bibnamefont {Prunet}},\ }\href
  {\doibase 10.1088/1475-7516/2013/02/001} {\bibfield  {journal} {\bibinfo
  {journal} {JCAP}\ }\textbf {\bibinfo {volume} {1302}},\ \bibinfo {pages}
  {001} (\bibinfo {year} {2013})},\ \Eprint {http://arxiv.org/abs/1210.7183}
  {arXiv:1210.7183 [astro-ph.CO]} \BibitemShut {NoStop}%
\bibitem [{\citenamefont {Lesgourgues}(2011)}]{Lesgourgues:2011re}%
  \BibitemOpen
  \bibfield  {author} {\bibinfo {author} {\bibfnamefont {J.}~\bibnamefont
  {Lesgourgues}},\ }\href@noop {} {\  (\bibinfo {year} {2011})},\ \Eprint
  {http://arxiv.org/abs/1104.2932} {arXiv:1104.2932 [astro-ph.IM]} \BibitemShut
  {NoStop}%
\bibitem [{\citenamefont {Blas}\ \emph {et~al.}(2011)\citenamefont {Blas},
  \citenamefont {Lesgourgues},\ and\ \citenamefont {Tram}}]{Blas:2011rf}%
  \BibitemOpen
  \bibfield  {author} {\bibinfo {author} {\bibfnamefont {D.}~\bibnamefont
  {Blas}}, \bibinfo {author} {\bibfnamefont {J.}~\bibnamefont {Lesgourgues}}, \
  and\ \bibinfo {author} {\bibfnamefont {T.}~\bibnamefont {Tram}},\ }\href
  {\doibase 10.1088/1475-7516/2011/07/034} {\bibfield  {journal} {\bibinfo
  {journal} {JCAP}\ }\textbf {\bibinfo {volume} {1107}},\ \bibinfo {pages}
  {034} (\bibinfo {year} {2011})},\ \Eprint {http://arxiv.org/abs/1104.2933}
  {arXiv:1104.2933 [astro-ph.CO]} \BibitemShut {NoStop}%
\bibitem [{\citenamefont {Ross}\ \emph
  {et~al.}(2015{\natexlab{b}})\citenamefont {Ross} \emph
  {et~al.}}]{Ross:2014mgs}%
  \BibitemOpen
  \bibfield  {author} {\bibinfo {author} {\bibfnamefont {J.~A.}\ \bibnamefont
  {Ross}} \emph {et~al.},\ }\href {\doibase 10.1093/mnras/stv154} {\bibfield
  {journal} {\bibinfo  {journal} {Mon. Not. Roy. Astron. Soc.}\ }\textbf
  {\bibinfo {volume} {449}},\ \bibinfo {pages} {835} (\bibinfo {year}
  {2015}{\natexlab{b}})},\ \Eprint {http://arxiv.org/abs/1409.3242}
  {arXiv:1409.3242 [astro-ph.CO]} \BibitemShut {NoStop}%
\bibitem [{\citenamefont {Poulin}\ \emph {et~al.}(2016)\citenamefont {Poulin},
  \citenamefont {Serpico},\ and\ \citenamefont {Lesgourgues}}]{Poulin:2016nat}%
  \BibitemOpen
  \bibfield  {author} {\bibinfo {author} {\bibfnamefont {V.}~\bibnamefont
  {Poulin}}, \bibinfo {author} {\bibfnamefont {P.~D.}\ \bibnamefont {Serpico}},
  \ and\ \bibinfo {author} {\bibfnamefont {J.}~\bibnamefont {Lesgourgues}},\
  }\href {\doibase 10.1088/1475-7516/2016/08/036} {\bibfield  {journal}
  {\bibinfo  {journal} {JCAP}\ }\textbf {\bibinfo {volume} {1608}},\ \bibinfo
  {pages} {036} (\bibinfo {year} {2016})},\ \Eprint
  {http://arxiv.org/abs/1606.02073} {arXiv:1606.02073 [astro-ph.CO]}
  \BibitemShut {NoStop}%
\end{thebibliography}%

\end{document}